\documentstyle[prd,aps,eqsecnum,epsf]{revtex}

\begin{document}

\title{The Chern-Simons state for the non-diagonal Bianchi IX model}
\author{Robert Paternoga and Robert Graham}
\address{Fachbereich Physik, Universit\"at-Gesamthochschule
Essen,
 45117 Essen, Germany}
\maketitle


\begin{abstract}
The Bianchi IX mixmaster model is quantized in its non-diagonal form,
imposing spatial diffeomorphism, time reparametrization and Lorentz
invariance  as constraints on physical state vectors {\em before} gauge-fixing.
The result turns out to be different from quantizing the diagonal model
obtained by gauge-fixing already on the classical level.
For the non-diagonal model a
generalized 9-dimensional Fourier transformation over a suitably chosen
manifold connects  the representations in metric variables and in
Ashtekar variables. A space of five states in the metric representation
is generated 
from the single physical Chern-Simons state in Ashtekar variables by 
choosing five different integration manifolds, which cannot 
be deformed into each other. For the case of a positive cosmological
constant $\Lambda$ we extend our previous study of these five states
for the diagonal
Bianchi IX model to the non-diagonal case. It is shown that additional 
discrete (permutation) symmetries of physical states arise in the quantization
of the {\it non-diagonal} model, which are satisfied by two of the five states
connected to the Chern-Simons state. These have the characteristics of a wormhole
groundstate and a Hartle-Hawking `no-boundary'
state, respectively. We also exhibit a
special gauge-fixing of the time reparametrization invariance of
the quantized system and
define an associated manifestly positive scalar product. Then the wormhole
ground state is left as the only normalizable physical state connected
to the Chern-Simons
state.
\end{abstract}

\section{Introduction}
Quantum general relativity has advanced over the last decade at
a remarkable
and accelerating pace. The introduction of Ashtekar's new
variables \cite{4,5,6} to replace the
earlier metric representation soon afterwards led to the discovery of a
formally exact physical state of the canonically quantized theory with
non-vanishing cosmological constant, the Chern-Simons state \cite{9,10}.
Then
the introduction of the loop representation \cite{loop} permitted to reexpress
the Chern-Simons state as a topological invariant of framed loops on 3-space
\cite{Wi}, the
Kauffman bracket \cite{Ka}. It also led to the discovery of further
physical states
lacking, however, 
one important general property of the Chern-Simons
state, namely a well-defined non-degenerate space-time in the 
classical limit.
The Chern-Simons state semiclassically describes a deSitter (or
anti-deSitter) space-time for positive (negative) cosmological constant,
respectively \cite{9}. Subsequent important advances were the introduction of
spin-networks \cite{RovSmol}
and quantum spin-networks \cite{MajSmol} as a discretized
description
of 3-space in which areas and volumes are quantized in Planck units and which
furnish yet another representation of physical states.

The choice of different variables and representations has therefore played,
and continues to play, a crucial role in the development of the theory.
It is not always clear, however, to which extent the different representations
are equivalent to each other. This
question is particularly relevant for the connection between the metric
representation and the representation in Ashtekar variables. In fact it was
shown in \cite{Mat} that the two representations are, in general, not
equivalent. One may therefore wonder: Has the Chern-Simons state, the
only known physical state with
a well-defined classical limit,  a  counterpart
in the metric representation and is it unique? In general, the answer
to this question
is still unknown.

Recently we examined this question for a spatially homogeneous
mini-superspace model of Bianchi type IX with diagonal metric tensor in the
case of positive \cite{0} and negative \cite{0-} cosmological constant $\Lambda$.
A {\it diagonal} form of the metric tensor can be assumed in the
{\it classical} theory without restriction of generality, because it is a
permissible gauge-fixing condition for the remnant of the spatial
diffeomorphism  group in the Bianchi type IX model.
Taking also the matrix of
Ashtekar variables as diagonal amounts in addition to a gauge-fixing of the
Lorentz gauge group. Quantizing such a diagonal model
therefore means to apply
gauge-fixing of the diffeomorphism group and the Lorentz group {\it before}
quantization. The result of our study of the diagonal Bianchi type IX model
was that actually five distinct physical states in the metric representation
are generated by transforming the Chern-Simons state from Ashtekar variables
to metric variables. This change of representation  takes
the form of a
generalized multidimensional Fourier
transformation in the  space spanned by the complex Ashtekar variables
along arbitrary paths or, more precisely,
integration manifolds with boundaries pinned
by the condition that partial
integration without boundary terms must be allowed. 

In the present work we take our investigation of this basic question a step
further and examine the metric representations of the Chern-Simons state for
the {\it non-diagonal} Bianchi type IX model. Why is this step interesting,
and why can't the results of our study of the diagonal case not just be
taken over?
Quantizing the non-diagonal model amounts to interchanging the steps of
gauge-fixing and quantizing: the gauge-fixing of the spatial diffeomorphism
group and Lorentz group is now done {\it after} the quantization. The result
is, in general, not the same as before the exchange of these steps. In
fact, in section III~A we give a very simple example from quantum
mechanics which shows that gauge-fixing {\it after} quantization is
preferable, in general, because it takes all symmetries into account in
the quantization process in a natural manner. The example also makes clear
that the result, in general, differs from that of {\it first} gauge-fixing
{\it then} quantizing by (i) quantum corrections in the Hamiltonian,
(ii) weight-functions in the naturally defined scalar product and
(iii) additional discrete symmetry requirements to be satisfied by the
solutions.

We shall see that all three differences also show up in our study of the
non-diagonal Bianchi type IX model. 
In fact the noncommutativity of gauge-fixing and quantization in the Bianchi
type IX model was previously discussed by Major and Smolin within the
framework of path integral quantization \cite{Maj1,Maj2}. Our investigation
here differs by the use of canonical quantization and, in particular,
by studying a special quantum state, the Chern-Simons state.
Comparing the two ways of quantization we find that 
on the leading semiclassical level the
solutions for the non-diagonal model are the same as in the diagonal case.
However, in the next to leading semiclassical order they already differ due
to the quantum corrections in the Hamiltonian and due to changes in the
naturally defined scalar product. Even more importantly, the additional
discrete symmetry requirements are only met by two of the five linearly independent
states, leaving just two physical states in the metric representation of the
non-diagonal model which are generated by the Chern-Simons state, a
generalized wormhole state and a Hartle-Hawking state \cite{16,19}.
It is remarkable that states of both kinds are related to the
Chern-Simons state and that the two different semiclassical boundary
conditions singling out either one of them can still be satisfied at this
stage.

In a final step we also gauge-fix (after quantization) the time
reparametrization invariance and introduce a manifestly positive scalar
product on the space of physical states. As will be shown in a separate paper
only the generalized wormhole state is normalizable in this scalar product.

The rest of this paper is organized as follows:
In section II we define our notation and set up the metric representation of
the constraint equations of the non-diagonal model (II~A), extract a
well-known exact solution for $\Lambda=0$, the wormhole state (II~B), and give
the representation of the constraint equations in Ashtekar variables (II~C).
In section III we discuss in detail the differences between the diagonal
and the non-diagonal Bianchi type IX model. We first give a simple
quantum mechanical example, the harmonic oscillator in two dimensions with
a rotational gauge symmetry (III~A). Then the corresponding comparison
for the Bianchi type IX model is given (III~B). We also mention here briefly
a gauge-fixing of the time reparametrization symmetry and an accompanying
physical inner product on the space of physical states (III~C) but a more
detailed presentation of this point is beyond the scope of the present
paper and will be given separately \cite{fut}. In section IV the transformation
of the Chern-Simons state of the non-diagonal Bianchi type IX model from the
Ashtekar representation (IV~A) to the metric representation (IV~B) is given,
leading to a general integral representation over a suitably chosen
3-dimensional manifold $\Sigma^3$. In section V various asymptotic forms of
this integral representation are evaluated. The limits considered are
{\it first} $\hbar\to 0$ (V~A) {\it then} either $\Lambda\to 0$ (V~A1) or
$\Lambda\,a^2 \to 0$ (V~A2), where $a$ is the geometrical mean of the scale
parameters, and $\Lambda \to 0$ (V~B) without necessarily taking a second limit.
In section VI we exhibit for the general case, i.e. without taking asymptotic
limits, five possible and topologically distinct choices of the integration
manifolds $\Sigma^3$ leading to five exact solutions of the constraint equations
in the metric representation. We also discuss their relation with the
asymptotic results of section V and their normalizability with respect to
the inner product of section III~C. Our conclusions are then summarized
in section VII.
Three appendices deal with certain technical details. The limit $\Lambda\to0$,
the results of which 
are discussed in section V~B, is quite subtle and therefore done
in some detail (A). It leads to a nice integral representation of the vacuum
(i.e. $\Lambda=0$) solutions (B). Finally we check certain required nontrivial
continuity and differentiability properties of the integrand of the integral
representation of section IV~B on the integration manifolds $\Sigma^3$ (C).

\section{The quantum constraint equations}

In this section we shall set up our notation and give a brief
derivation of the quantum Einstein equations 
for the homogeneous Bianchi IX model. While the classical constraint
equations for this model are determined uniquely, the quantum operators
associated with these constraints suffer from the well known ambiguity
of the factor ordering, in particular in the 
Hamiltonian constraint. The main purpose of
the following will be the motivation of a special choice of
factor ordering. Technical details of the derivation will
be summarized rather briefly.

\subsection{Metric representation of the constraint equations}

Let us start with the Einstein Hilbert action for a gravitational field
with a cosmological constant $\Lambda$:

\begin{equation}\label{1}
{\cal S}_{\mbox{{\tiny $E\!H$}}}\ [g_{\mu \nu}]\ =\ 
\int\limits_{{\cal M}} \mbox{d} ^{4} x \sqrt{-g}\ \Bigl(\ ^{4}{\cal R}-2
\Lambda \Bigr )
\end{equation}
\\
Here ${\cal M}$ is the space-time manifold, {\boldmath $g$}
$=(g_{\mu \nu})$ the 4-metric on ${\cal M}$, $g=\det\,(g_{\mu \nu})$
and $^{4}{\cal R}$ the curvature scalar of the 4-metric. The common 
prefactor $\frac{1}{16 \pi G}$ of eq. (\ref{1}) has been avoided by
picking units with $G=(16 \pi )^{-1}$. As is well known, the ADM
space-time split and a subsequent Legendre transform with respect to
$\dot h_{i j}$ yield the following equivalent expression  for
the Einstein Hilbert action \cite{17}:

\begin{equation}\label{2.1}
{\cal S}_{\mbox{{\tiny $E\!H$}}} =
\int \mbox{d} t \int \mbox{d}^3 x \left (
\tilde \pi^{i j} \, \dot h_{i j} - N \, {\cal H}_G - N_i \, {\cal H}^i
\right ) \ \ ,
\end{equation}
\begin{equation}\label{2.2}
\mbox{with}\ \ \ 
{\cal H}_G = G_{i j k l} \,\tilde \pi^{i j} \,\tilde \pi^{k l}
- \sqrt{h} \, \left ( \, ^3{\cal R} -2 \Lambda \right )\ \ \ \ ,
\ \ \ {\cal H}^i = -2\,\tilde \pi^{i j}_{\ \ |j}\ \ ,
\end{equation}  
\begin{equation}\label{2.3}
\mbox{where}\ \ \ 
G_{i j k l}:=\frac{1}{2 \sqrt{h}}\, \big (
h_{i k} \,h_{j l}+h_{i l} \, h_{j k}-h_{i j}\, h_{k l} \big )\ \ .
\end{equation}  
\\
In the transition from (\ref{1}) to (\ref{2.1}) a surface term has
been omitted, since it will have no effect on the constraint equations.
$N$ and $N_i$ are the lapse function and the shift vector, respectively,
and the 3-metric $(h_{i j})$ of the spatial manifold $t=const.$ 
is used to raise and lower the spatial indices $i, j, k, \dots$.
A stroke denotes covariant derivatives with respect to the 3-metric,
$^3{\cal R}$ is the curvature scalar of the spatial manifold and $h$
denotes the determinant of the 3-metric.

Let us now consider a {\em homogeneous} 3-manifold of one of the Bianchi
types; then there exists an invariant basis of one-forms
{\boldmath$\omega$}$^p=
\omega^p_{\ i}\,(x)\,\mbox{d}x^i$, such that any homogeneous tensor field
on the manifold has spatially constant components when expanded in this
basis \cite{18,18+}. In particular, we have 

\begin{equation}\label{3}
\mbox{d{\boldmath$\omega$}}^p = \frac{1}{2}\,m^{p q}\,\varepsilon_{q r s}
\,\mbox{{\boldmath$\omega$}}^r \wedge
\,\mbox{{\boldmath$\omega$}}^s\ \ ,
\end{equation}
\\
with a constant structure matrix $m^{p q}$. In the following we shall
be interested in the Bianchi type IX case, where the structure matrix
is of the simple form $m^{p q} =\delta^{p q}$. 
If all the tensor fields
occuring in (\ref{2.1}) are expanded in the invariant basis, we arrive
at the following expression for the Einstein Hilbert action\footnote{
The following expressions are valid not only for the Bianchi IX model
with $m^{p q}=\delta^{p q}$, but even for a general Bianchi A model
characterized by $m^{p q}=m^{q p}$.}:

\begin{equation}\label{4.1}
{\cal S}_{\mbox{{\tiny $E\!H$}}} =
\int \mbox{d} t\,{\cal L}_{\mbox{{\tiny $E\!H$}}}=
\int \mbox{d} t \int \mbox{d}^3 x \ \omega\ \left (
\tilde \pi^{p q} \, \dot h_{p q} - N \, {\cal H}_0 - N^p \, {\cal H}_p
\right ) \ \ ,
\end{equation}
\begin{equation}\label{4.2}
{\cal H}_0 = G_{p q r s} \,\tilde \pi^{p q} \,\tilde \pi^{r s}
- \sqrt{h} \, \left ( \, ^3{\cal R} -2 \Lambda \right )\ \ \ \ ,
\ \ \ {\cal H}_p = 2\,\varepsilon_{p r q}\,m^{q n}\,h_{n s}\,
\tilde \pi^{s r}\ \ ,
\end{equation}  
\begin{equation}\label{4.3}
\mbox{where}\ \ \ 
h \,^3{\cal R} =\frac{1}{2}\, \left (m^{p s}\,h_{s p} \right )^2 -
m^{p s}\,h_{s q}\,m^{q r}\,h_{r p}\ \ .
\end{equation}  
\\
Here $\omega=\det \,(\omega^p_{\ i})$ 
in (\ref{4.1}) contains the only spatial dependence of
${\cal S}_{\mbox{{\tiny $E\!H$}}}$,
which therefore can be integrated out explicitely. Afterwards, 
the Lagrangian equations with respect to $N$ and $N^p$ imply
the following set of first class constraints

\begin{equation}\label{5}
{\cal H}_0 = 0\ \ \ , \ \ \ 
{\cal H}_p = 0\ \ \ ,
\end{equation}
\\
with ${\cal H}_0$ and ${\cal H}_p$ being phase space functions of the
fundamental variables $h_{p q}$ and the momenta
$\tilde \pi^{p q}$ only.\footnote{
According to (\ref{4.1}) $\tilde \pi^{p q}$ are {\em not} the canonically
conjugate momenta to $h_{p q}$; they differ from them by a rescaling
factor $V:=\int \mbox{d}^3 x\,\omega$.
Consequently, the canonical Poisson-brackets
read 

\begin{displaymath}
\{ \tilde \pi^{p q}, h_{r s}\} =-\frac{1}{V}\,\delta^{p q}_{r s} \ \ ,
\ \ \mbox{with}\ \ 
\delta^{p q}_{r s}:=\frac{1}{2} (\delta^p_r\,\delta^q_s+
\delta^p_s\,\delta^q_r),
\end{displaymath}
\\
and the additional factor $\frac{1}{V}$ should be carried over to the 
canonical commutation relations of the quantized theory. There, it can
be eliminated by a rescaling of Planck's constant $\hbar$. Since we
are mainly interested in the quantized theory, we will use this freedom
to set $V=1$ in the following.}   
If we make use of the remarkable identity

\begin{equation}\label{6}
\sqrt{h} \ ^3{\cal R} =- m^{p q}\,G_{p  q r s}\,m^{rs}\ \ ,
\end{equation}
\\
which is valid for the general Bianchi type A case,
the Hamiltonian constraint ${\cal H}_0$ may be written in one of the
two following forms:

\begin{equation}\label{7}
{\cal H}_0 = \big [\tilde \pi^{p q}\pm i\,m^{p q} \big ]\,G_{p  q r s}\,
\big [\tilde \pi^{r s} \mp i\,m^{r s} \big ]+2 \Lambda\,\sqrt{h}\ \ .
\end{equation}
\\
Let us first restrict ourselves to the kinetic part (which
becomes the full Hamiltonian constraint for the Bianchi type I model)

\begin{equation}\label{8}
T = \tilde \pi^{p q}\,G_{p  q r s}\,\tilde \pi^{r s}\ \ ,
\end{equation}
\\
describing a particle moving freely on a six-dimensional manifold
with coordinates $h_{p q}$ and the (indefinite) supermetric 
$G^{p q r s}$, which is the inverse of $G_{p q r s}$.\footnote{
More precisely, we require $G^{p q m n}\,G_{m n r s} =\delta^{p q}_{r s}$.} 
To quantize such a system, we may employ new coordinates $h'_{p q}$
of the minisuperspace such that, at least locally, the supermetric
takes a diagonal form, i.e.

\begin{equation}\label{9}
T = {\tilde \pi}^{p q}\,'\,\eta_{p  q r s}\,{\tilde \pi}^{r s}\,'\ \ \ ,
\ \ \ \mbox{with}
\ \ \ \eta_{p  q r s} :=\biggl\{
\begin{array}{ccl}
-1&\ \ ,&\mbox{if} \ \ p=q=r=s=1\ ,\\
\frac{1}{2} (\delta_{p r}\,\delta_{q s}+\delta_{p s}\,\delta_{q r})
&\ \ ,&\mbox{otherwise}\ \ .
\end{array}
\end{equation}
\\
In these `free falling' superspace coordinates the associated quantum operator
is expected to be

\begin{equation}\label{10}
\hat T = - \hbar^2 \ \frac{\partial}{\partial h'_{p q}}\,
\eta_{p  q r s}\,
\frac{\partial}{\partial h'_{r s}}\ \ ,
\end{equation}
\\
and, transforming back to the $h_{p q}$-coordinates, we arrive at

\begin{equation}\label{11}
\hat T = - \hbar^2 \ \frac{1}{\sqrt{G}}\ 
\frac{\partial}{\partial h_{p q}}\,
\sqrt{G}\,G_{p  q r s}\,
\frac{\partial}{\partial h_{r s}}\ \ ,
\end{equation}
\\
which is easily recognized as the invariant Laplace Beltrami operator 
on minisuperspace. Here $G$ is the absolute value of the determinant of the supermetric
$G^{p q r s}$, which can be shown to be proportional to $h^{-1}$.

These considerations suggest the following factor
ordering for the quantum version of the Hamiltonian constraint (\ref{7}) 

\begin{eqnarray}\label{12}
\hat {\cal H}_0&=&
\frac{1}{\sqrt{G}}\ \left [ 
-i\,\hbar\,\frac{\partial}{\partial h_{p q}}+i\,m^{p q} \right ]\,
\sqrt{G}\,G_{p  q r s}\,
\left [-i\,\hbar\,\frac{\partial}{\partial h_{r s}}-i\,m^{r s} \right ]
+2 \Lambda\,\sqrt{h}\nonumber\\[.5 cm]
&=&
- \hbar^2 \ \sqrt{h}\ 
\frac{\partial}{\partial h_{p q}}\,
\frac{1}{\sqrt{h}}\,G_{p  q r s}\,
\frac{\partial}{\partial h_{r s}}-\hbar\,\sqrt{h}\,
\frac{\partial}{\partial h_{p q}}\,\left (
\frac{1}{\sqrt{h}}\,G_{p q r s}\,m^{r s} \right )- \sqrt{h}
\, \left ( \, ^3{\cal R} -2 \Lambda \right )\ \ ,
\end{eqnarray}
\\
where we have chosen the signs of $i\, m^{p q}$ in the first line to
make the special physical state defined below in
(\ref{18}) become an exponentially
decaying solution.
We end up with a Wheeler DeWitt operator, which, apart from a special
factor ordering in the kinetic term, contains a quantum correction
to the potential, due to the action of the derivative operators on the
supermetric. Such a term is well-known from the diagonal model, cf.
\cite{0}.
To put the diffeomorphism constraint in more concrete terms, we should
be further interested in a $h_{p q}$-representation of 
$\tilde \pi^{p q}$, obeying the canonical commutation relations

\begin{equation}\label{13}
\left [ \tilde \pi^{p q}, h_{r s} \right ]= - i\, \hbar\,\delta^{p q}_{r s}
\ \ .
\end{equation}
\\
Moreover, we wish $\tilde \pi^{p q}$ to be a hermitian operator
with respect to the natural auxiliary inner product on the extanded
Hilbert space, in which the Wheeler DeWitt operator (\ref{12}) is
hermitian,

\begin{equation}\label{ip}
\langle \Psi | \Phi \rangle = \int \mbox{d}^9 h_{p q}
\ \prod_{p>q}\,\delta (h_{p q}-h_{q p})\ \sqrt{G}\  \Psi^* (h_{r s}) \,\Phi(h_{r s})
\ \ .
\end{equation}
\\
Here the components of $h_{p q}$ range over all real values which are 
consistent with $(h_{p q})$ having positive eigenvalues.\footnote{
The so-defined integration regime for the auxiliary inner product
has a non-trivial boundary at $h=0$, where at least one of the
three eigenvalues of $(h_{p q})$ vanishes. Consequently, we have to
restrict ourselves to states $\Psi$, which vanish for $h \to 0$
in a suitable manner to assure hermiticity of the differential operators
occuring in (\ref{12}).}
As is known from the quantization on a curved manifold, the 
$h_{p q}$-representation of the momentum operator has to be corrected with a
relative weight factor $G^{1/4}$, leading to

\begin{equation}\label{14}
\tilde \pi^{p q}=-i\, \hbar\,G^{-1/4}\,
\frac{1}{2}\left (
\frac{\partial}{\partial h_{p q}}+\frac{\partial}{\partial h_{q p}}
\right )\,G^{1/4}=
-i\, \hbar\,h^{1/4}\,
\frac{\partial}{\partial h_{(p q)}}\,h^{-1/4}\ \ .
\end{equation}  
\\
Using this $h_{p q}$-representation, we may rewrite $\hat {\cal H}_0$
and $\hat {\cal H}_p$ in the form

\begin{equation}\label{15}
\hat {\cal H}_0 = (a^{p q})^{\dag}\, G_{p q r s}\, a^{r s}
+2\,\Lambda\sqrt{h} = 
\hat {\cal H}_0^{\dag}\ \ \ \mbox{with}\ \ \ 
a^{p q}:=h^{-1/4}\,
\left [\tilde \pi^{p q}-i\,m^{ p q} \right ] 
\,h^{1/4}\ \ ,
\end{equation}  
\begin{equation}\label{15.1}
\hat {\cal H}_p =
2\,\varepsilon_{p r q}\,m^{q n}\,h_{n s}\,\tilde \pi^{s r}
=
2\,\varepsilon_{p r q}\,m^{q n}\,h_{n s}\,a^{s r}
=
-2\, i\, \hbar\,\varepsilon_{p r q}\,m^{q n}\,h_{n s}\,
\frac{\partial}{\partial h_{(s r)}}=\hat {\cal H}_p^{\dag}\ \ .
\end{equation}
\\
That the diffeomorphism constraints can indeed be written in the form
of (\ref{15.1}) is checked easily by using the identity

\begin{equation}\label{16}
h^{1/4}\,\frac{\partial}{\partial h_{(p q)}}\,
h^{-1/4}=-\frac{1}{4}\,h^{p q}\ \ ,
\end{equation}
\\
which implies that there is no additional contribution to $\hat {\cal H}_p$
arising from the determinant factor, as one might expect.
With (\ref{15}) and (\ref{15.1})
we have now nice, self-adjoint operators, which, moreover, form a
closed algebra, as we will see in eq. (\ref{22}) below.

\subsection{The wormhole state}

As is immediately seen from (\ref{12}), the wavefunction

\begin{equation}\label{17}
\Psi_{\mbox{{\tiny $W\!H$}}}:=
\exp\,\left [ -\frac{1}{\hbar}\,m^{p q}\,h_{p q} \right ]
=:\exp\, \left [-\frac{\Phi}{\hbar} \right]
\end{equation}
\\
is a solution to the Hamiltonian constraint for $\Lambda =0$, which
moreover satisfies $a^{p q}\,\Psi_{\mbox{{\tiny $W\!H$}}}
=0$ and therefore solves the diffeomorphism constraint (\ref{15.1})
as well. For the Bianchi IX model, (\ref{17}) may also be written in
the form 

\begin{equation}\label{18}
\Psi_{\mbox{{\tiny $W\!H$}}}=
\exp\,\left [ -\frac{1}{\hbar}\,(\lambda_1+\lambda_2+\lambda_3) \right ]
\ \ ,
\end{equation}
\\
where $\lambda_p$ are the eigenvalues of the 3-metric $h_{p q}$, and
this state is known as the `wormhole' state from the diagonal
model.\footnote{This name derives from the fact that $\Phi$ solves the
Euclideanized Hamilton Jacobi equation with Euclidean four-geometry
at large scale parameter. We shall use this name in the following
to refer to the states (\ref{17}) or (\ref{18}) for $\Lambda =0$ or its
generalization for $\Lambda \not=0$, without entering a discussion
of wormholes, however.}
The fact that it will occur as a prefactor to all further
wavefunctions discussed
within this paper suggests to perform the following
similarity transformation:

\begin{equation}\label{26}
\Psi=e^{-\Phi/\hbar}\,\Psi'\ \ ,
\ \ \hat{\cal H}_p=e^{-\Phi/\hbar}\,\hat{\cal H}'_p\,
e^{\Phi/\hbar}\ \ ,\ \  
\hat{\cal H}_0=e^{-\Phi/\hbar}\,\hat{\cal H}'_0\,
e^{\Phi/\hbar}\ \ .
\end{equation}
\\
In this new representation, the transformed operators take the form

\begin{equation}\label{25.1}
\hat{\cal H}'_p 
=
-2 i\, \hbar\,\varepsilon_{p r q}\,m^{q n}\,h_{n s}\,
\frac{\partial}{\partial h_{(s r)}}
\ \ ,
\end{equation}
\begin{equation}\label{25.2}
\hat{\cal H}'_0=
\frac{1}{\sqrt{G}}\ \left [ 
i\,\hbar\,\frac{\partial}{\partial h_{p q}}-2 i\,m^{p q} \right ]\,
\sqrt{G}\,G_{p  q r s}\,
i\,\hbar\,\frac{\partial}{\partial h_{r s}}
+2 \Lambda\,\sqrt{h}\ \ , 
\end{equation}
\\
which will  be recovered from a very different approach in the next section.

\subsection{Representation of the constraint equations in Ashtekar
variables}

In this section we want to derive the so-called Ashtekar
representation \cite{4,5}
of the quantized non-diagonal Bianchi IX
model using the inverse densitized triad of the 3-metric $h_{p q}$
and complexified canonically conjugate variables thereof.
The first step in this direction has already been performed
in the last section by splitting off the wormhole state (\ref{17}).
What remains to be done now is to introduce the inverse,
densitized triad of the 3-metric
$h_{p q}$, defined via

\begin{equation}\label{triad}
h\! \cdot \! h^{p q}=\tilde e^{p}_{\ a}\! \cdot \tilde e^{q}_{\ a}\ \ .
\end{equation}
\\
Here and in the following  $a,b,c,\dots$ are flat, internal
indices running from one to
three; they are raised and lowered with the flat metric $\delta_{a b}$
and will therefore always be chosen to be lower indices without any
restriction. The introduction of a triad has the great advantage that
the three metric defined via (\ref{triad}) is automatically positive
definite, at least as long as the triad is real-valued, and this
is favourable for a definition of an inner product on the space
of wavefunctions. 
However, as is well-known, we gain three additional degrees of freedom by
introducing such a triad, corresponding to the three possible rotations
in the flat local tangent space. These redundant 
rotational degrees of freedom are accompanied by three additional first
class constraints
referred to
as the Gau{\ss}-constraints. 

To construct the generators of these constraints 
we observe that eq. (\ref{triad}) is invariant under rotations of $\tilde
e^{p}_{\ a}$ in the local tangent space, generated by 

\begin{equation}\label{2.34}
J'_a=i\,\varepsilon_{a b c}\,{\cal A}_{p b}\,\tilde e^{p}_{\ c}
\ \ \ \mbox{with}\ \ \ {\cal A}_{q a}:=-\frac{\hbar}{2}\,\frac{\partial}
{\partial\,\tilde e^{q}_{\ a}}\ \ .
\end{equation}
\\
The $J'_{a}$ satisfy the angular momentum algebra

\begin{equation}\label{2.35}
[J'_{a},J'_{b}]=i\,\hbar\,\varepsilon_{a b c}\,J'_{c}
\end{equation}
\\
and commute with $\hat{\cal H}'_{p}$ and $\hat{\cal H}'_{0}$. They are the
generators of the Gau{\ss} constraints. The general solution of the
Gau{\ss} constraints $J'_{a}\,\psi'=0$ is obviously given by wavefunctions
of the form 

\begin{equation}\label{23}
\psi'(\tilde e^p_{\ a})=\Psi'(h_{p q})\ \ ,
\end{equation}
\\
where $h_{p q}$ is a function of the $\tilde e^{p}_{\ a}$ via (\ref{triad}).
Acting on solutions of the Gau{\ss} constraints the operator
$\partial / \partial \tilde e^{p}_{\ a}$ can be written as

\begin{equation}\label{24}
{\cal A}_{p a}=-\frac{\hbar}{2}\,
\frac{\partial}{\partial \tilde e^p_{\ a}}=
\frac{\hbar}{e}\,
G_{m n p q}\  \tilde e^q_{\ a}\,\frac{\partial}{\partial h_{m n}}\ \ ,
\end{equation}
\\
where $e$ is the square root of the
determinant of $\tilde e^{p}_{\ a}$. This permits us
to rewrite the constraint operators (\ref{25.1}) and (\ref{25.2}) in terms of
the operators $\tilde e^p_{\ a}$ and ${\cal A}_{p a}$ as

\begin{equation}\label{20.2}
{\cal H}'_p:=
2 i\,\varepsilon_{p q r}\,m^{r s}\,
\tilde e^q_{\ a}\,{\cal A}_{s a}\ \ ,
\end{equation}
\begin{equation}\label{20.3}
{\cal H}'_0:=e^{-1}\,
\varepsilon_{a b c}\,\varepsilon_{p q r}\,
\tilde e^p_{\ a}\tilde e^q_{\ b}\,{\cal Q}^r_{\ c}\ \ ,
\end{equation}
\\
with the operators 

\begin{equation}\label{19}
{\cal Q}^p_{\ a}:={\cal F}^p_{\ a}+\frac{\Lambda}{3}\,\tilde e^p_{\ a}
:=m^{p q}\,{\cal A}_{q a}+\frac{1}{2}\,\varepsilon^{p q r}\,
\varepsilon_{a b c}\,{\cal A}_{q b}\,{\cal A}_{r c}
+\frac{\Lambda}{3}\,\tilde e^p_{\ a}\ \ .
\end{equation}
\\
The operators ${\cal Q}^p_{\ a}$ are very convenient and can also be
used to reexpress the $J'_a$ and ${\cal H}'_p$ as 

\begin{equation}\label{21.1}
J'_a:=\frac{3 i}{\Lambda}\,
\varepsilon_{a b c}\,{\cal A}_{p b}\,{\cal Q}^p_{\ c}
\ \ ,
\end{equation}
\begin{equation}\label{21.2}
{\cal H}'_p=
2 i\,\left( \varepsilon_{p q r}\,\tilde e^q_{\ a}\,
{\cal Q}^r_{\ a}-{\cal A}_{p a}\,J'_a\ \right )\ .
\end{equation}
\\
By construction ${\cal H}'_p$ and ${\cal H}'_0$ coincide with 
$\hat {\cal H}'_p$ and $\hat {\cal H}'_0$ defined in (\ref{25.1})
and (\ref{25.2}) when acting on the invariant 
subspace spanned by the solutions of the Gau{\ss} constraints (but they
extend these operators also to non physical states outside this 
subspace, which are of no interest to us, however). The commutation
relations of ${\cal H}'_p$ and ${\cal H}'_0$ are particularly easy
to evaluate in the representation (\ref{20.2}), (\ref{20.3}) 

\begin{equation}\label{22}
\big[{\cal H}'_p,{\cal H}'_q\, \big]=i\,
\hbar\,\varepsilon_{p q r}\,m^{r s}
\,{\cal H}'_s\ \ ,\ \ 
\big[{\cal H}'_0,{\cal H}'_p\,\big]=0\ \ .
\end{equation}
\\
They imply the same commutator algebra for the
$\hat {\cal H}'_p$, $\hat {\cal H}'_0$, and, via the similarity
transformation (\ref{26}), also for the $\hat {\cal H}_p$, $\hat {\cal H}_0$.
 
Summarizing our results so far , we have shown that, assuming that
$\psi'(\tilde e^p_{\ a})$
is a solution to the constraints (\ref{2.34}), (\ref{20.2}) and
(\ref{20.3}), the transformed wavefunction $\Psi(h_{p q})$, connected
with $\Psi'$ via (\ref{26}), is a solution to the constraints (\ref{15})
and (\ref{15.1}) in the metric representation. In particular, if we are able 
to solve the more restrictive but simpler set of equations

\begin{equation}\label{29}
{\cal Q}^p_{\ a}\,\psi'(\tilde e^p_{\ a})=0\ \ ,
\end{equation}
\\
it is clear from the definitions given in (\ref{20.3}),
(\ref{21.1}) and (\ref{21.2})
that we have also 
found a quantum state of the full, non-diagonal Bianchi IX model with
a non-vanishing cosmological constant in the metric representation. This
will only be a special class of solutions, however, because (\ref{29})
represents nine conditions, while the Hamiltonian, diffeomorphism and
Gau{\ss} constraints together constitute seven conditions only.

\section{Comparison between the diagonal and non-diagonal Bianchi IX
model}

We shall now adress the interesting question,
wether the non-diagonal Bianchi IX model presented in section II
and the diagonal Bianchi IX model discussed in \cite{0} have the same
physical content. On the classical level, it is of course unnecessary to
distinguish between these two models, because we  can use the gauge freedom 
of the diffeomorphism constraints to transform the non-diagonal Bianchi
IX model to the diagonal one. When discussing the {\em quantized} Bianchi IX 
model, most authors restricted themselves to the diagonal case, i.e.
they solved for the diffeomorphism constraints on the classical level,
and performed the canonical quantization procedure for the effective,
3 dimensional system. This approach 
immediately suggests itself, and one may hope
that a quantization of the full Bianchi IX model (on a 6-dimensional
configuration space) with diffeomorphism constraints imposed on the
quantum mechanical level should physically lead to the same results.

In the following, we shall first discuss a very simple example, which
immediately shows that this believe is in general not true; we will then
show in subsection B that the two quantization procedures used for
the diagonal and non-diagonal Bianchi IX model indeed differ 
drastically.

\subsection{An example: The 2 dimensional harmonic oscillator with $L=0$}

Let us consider a well known example, the 2 dimensional harmonic oscillator
with unit mass and unit frequency, but with the additional constraint
that the angular momentum should vanish:

\begin{eqnarray}\label{H1}
&{\cal H}={\cal H}_0+N\,L\ \ ,&\nonumber\\[.3 cm]
&{\cal H}_0=\frac{1}{2}\,(p_1^2+p_2^2)+\frac{1}{2}\,(q_1^2+q_2^2)\ \ \ ,\ \ \ 
L=q_1\,p_2-q_2\,p_1\ \ .&
\end{eqnarray}
\\
Here ${\cal H}_0$ is the Hamiltonian of the unconstrained system and
$N$ is a Lagrangian multiplier. As for the Bianchi type IX model, we have
the nice property that the constraint is given by a conserved quantity,
but $L$ generates {\em gauge} transformations only if we identify all the
directions, in which the effective one dimensional oscillator can move.

If we firstly quantize this system similarly to the diagonal Bianchi IX case,
we have to solve $L=0$ on the classical level, which is done by

\begin{equation}\label{H2}
q_2=0\ \ \Rightarrow \ \ p_2=\dot q_2=0\ \ .
\end{equation}
\\
In this gauge we arrive at the effective Hamiltonian

\begin{equation}\label{H3}
{\cal H}_{e\! f\! f}=
\frac{1}{2}\,p^2+\frac{1}{2}\,q^2\ \ , \ \ q \equiv q_1\ \ ,
\end{equation}
\\
what is easily quantized because it simply describes a one dimensional
harmonic oscillator:

\begin{equation}\label{H4}
\hat {\cal H}_{e\! f\! f}=-\frac{\hbar^2}{\! 2}\,\frac{\partial^2}{\partial q^2}+
\frac{1}{2}\,q^2\ \ ,\ \ 
\langle \Psi | \Phi \rangle =\int\limits_{-\infty}^{+\infty}
\mbox{d} q \, \Psi^*(q)\, \Phi(q)\ \ .
\end{equation}
\\
Let us now secondly proceed in analogy to the non-diagonal Bianchi IX
case. Then we have to quantize first, and obtain

\begin{equation}\label{H5}
\hat {\cal H}=\hat{\cal H}_0+N\,\hat L\ \ ,\ \ 
\langle \psi | \phi \rangle =\int_{
\mbox{{\boldmath$R$}}^2} \mbox{d}^2 q\ 
\psi^*(q_1,q_2) \,\phi(q_1,q_2)
\end{equation}
\begin{equation}\label{H5+}
\mbox{with}\ \ 
\hat {\cal H}_0=- \frac{\hbar^2}{\! 2} \, \left (
\frac{\partial^2}{\partial q_1^2}+\frac{\partial^2}{\partial q_2^2}
\right )+\frac{1}{2}\,(q_1^2+q_2^2)\ \ ,\ \ 
\hat L=-i\,\hbar\,\left (
q_1\,\frac{\partial}{\partial q_2}-q_2\,\frac{\partial}{\partial q_1}
\right )\ \ .
\end{equation}
\\
Solving $\hat L \, \psi=0$, the wavefunction must be of the form

\begin{equation}\label{H6}
\psi(q_1,q_2)=\Psi(q)\ \ ,\ \ q=\sqrt{q_1^2+q_2^2}\ \ .
\end{equation}
\\
However, to ensure that $\psi$ is a differentiable function with respect
to $q_1$ and $q_2$,
we must require $\Psi(q)$ to be an even function in $q$,

\begin{equation}\label{H7}
\Psi(q)=\Psi(-q)\ \ .
\end{equation}
\\
Furthermore, if we compute the action of $\hat {\cal H}$ on such a solution
$\Psi(q)$, we effectively have

\begin{equation}\label{H8}
\hat {\cal H}_{e\! f\! f}=- \frac{\hbar^2}{\! 2}\,\frac{1}{q}\,
\frac{\partial}{\partial q}\,q\,\frac{\partial}{\partial q}+\frac{1}{2}\,q^2
=- \frac{\hbar^2}{\! 2}\,\frac{\partial^2}{\partial q^2}
- \frac{\hbar^2}{2 \, q}\,\frac{\partial}{\partial q}+
\frac{1}{2}\,q^2\ \ ,
\end{equation}
\\
and the effective scalar product for solutions to the angular momentum
constraint becomes

\begin{equation}\label{H9}
\langle \psi | \phi \rangle =
\int_{\mbox{{\boldmath$R$}}^2} \mbox{d}^2 q\ 
\psi^*(q_1, q_2)\,\phi(q_1, q_2)
=2\,\pi\,\int\limits_{0}^{\infty} \mbox{d} q\,q\,\Psi^*(q)\,\Phi(q)
=\langle \Psi | \Phi \rangle_{e\! f\! f}\ \ .
\end{equation}
\\
As a result, we have three important differences between the quantization
procedures pointed out above:

\begin{itemize}

\item[(i)] The effective Hamilton operators (\ref{H4}) and (\ref{H8})
differ by a factor ordering term $- \frac{\hbar^2}{2\,q}\,
\frac{\partial}{\partial q}$, which becomes singular where the 
coordinate transformation from $q_1, q_2$ to polar coordinates
with radius $q$ is not invertible.

\item[(ii)] The scalar products (\ref{H4}) and (\ref{H9}) contain
different weight functions and different integration regimes.

\item [(iii)] For the second quantization procedure, we get a parity
requirement (\ref{H7}) being absent in the first case, because we originally
start with differentiable functions on a higher dimensional
configuration space.

\end{itemize}

\noindent
All three differences will now be recovered when comparing the
diagonal and non-diagonal Bianchi IX model.

\subsection{Three differences between the quantized diagonal and 
non-diagonal Bianchi IX model}

To compare our results of section II for the non-diagonal Bianchi
IX model to the diagonal case discussed in \cite{0}, let us try to
solve the diffeomorphism constraints (\ref{15.1}) on the quantum mechanical
level. Three special, regular solutions to these constraint equations
$\hat {\cal H}_p\,\Psi (h_{p q})=0$ are the invariants of the 3-metric,
which read

\begin{equation}\label{D1}
T:=\mbox{Tr}\,(h_{p q})\ \ ,\ \ Q:=\delta_{m n}\,
\varepsilon^{m p q}\,\varepsilon^{n r s}\,h_{p r}\,h_{q s}\ \ ,
\ \ h=\det\, (h_{p q})\ \ .
\end{equation}
\\
Therefore the general solution to $\hat {\cal H}_p\,\Psi=0$ is any function
of these three invariants,

\begin{equation}\label{D2}
\Psi(h_{p q})=\chi(T,Q,h)\ \ ,
\end{equation}
\\
where $\chi(T,Q,h)$ is a differentiable function with respect to its
three arguments $T,Q,h$. We may now express this general solution in terms
of the three eigenvalues $\lambda_p$ of the 3-metric, which, however,
are {\em not} $C^1(\mbox{{\boldmath$R$}}^6)$-functions of $h_{p q}$,
because of the cubic roots which are needed to express the $\lambda_p$
in terms of $h_{p q}$. Nevertheless, a function $\Psi(h_{p q})$ 
solving $\hat {\cal H}_p \, \Psi =0$ according to (\ref{D2}) actually {\em is}
$C^1(\mbox{{\boldmath$R$}}^6)$ with respect to $h_{p q}$, so

\begin{equation}\label{D3}
\psi (\lambda_p)=\Psi(h_{p q})=\chi(\lambda_1+\lambda_2+\lambda_3,
\,\lambda_1\,\lambda_2+\lambda_2\,\lambda_3+\lambda_3\,\lambda_1,
\,\lambda_1\,\lambda_2\,\lambda_3)
\end{equation}
\\
is a regular function in $h_{p q}$, too. We observe that, as a consequence,
$\psi(\lambda_p)$ is symmetric under arbitrary permutations of the
$\lambda_p$, while any wavefunction, which is not symmetric under these
permutations will not be differentiable with respect to $h_{p q}$. This
symmetry requirement is the analog to the parity requirement (\ref{H7}) of
our example in subsection A. 

It will be convenient to introduce new variables 

\begin{equation}\label{sig}
\sigma_p:=\frac{2}{\hbar}\,
\sqrt{\lambda_q \lambda_r} >0\ \ ,\ \ \mbox{with} \ \ \varepsilon_{p q r}=1
\ \ ,
\end{equation}
\\
instead of the eigenvalues $\lambda_p$. In the diagonal gauge
they play the role of the inverse
densitized triad. The inverse transformation
reads 

\begin{equation}\label{D5}
\lambda_p=\frac{\hbar}{2}\,\frac{\sigma_q\,\sigma_r}{\sigma_p}\ \ ,
\end{equation}
\\
and we arrive at the representation

\begin{equation}\label{D6}
\Psi(\sigma_p)=\psi(\lambda_p)=
\psi(\frac{\hbar}{2}\,\frac{\sigma_q\,\sigma_r}{\sigma_p})\ \ ,
\end{equation}
\\
so the wavefunction $\Psi(\sigma_p)$ is not only invariant under
arbitrary permutations of the $\sigma_p$, but in addition invariant
under reflexions $\sigma_p \rightarrow -\sigma_p,
\sigma_q \rightarrow -\sigma_q,\sigma_r \rightarrow \sigma_r$. These
are {\em necessary} symmetry requirements for the wavefunctions in the
$\sigma_p$-representation, which were absent in the diagonal case, cf.
\cite{0}.

The effective Hamiltonian constraint in the $\sigma_p$-representation
becomes

\begin{equation}\label{31.1}
{\cal H}'_0\, \Psi' \propto \, \sqrt{\sigma_1\,\sigma_2\,\sigma_3}\,
\left [
\frac{1}{\sigma_1}\,{\cal Q}_1+
\frac{1}{\sigma_2}\,{\cal Q}_2+
\frac{1}{\sigma_3}\,{\cal Q}_3
\right ]\,\Psi'=0\ \ ,
\end{equation}     
\begin{equation}\label{31.2}
\mbox{with}\ \ \ {\cal Q}_p:=
\partial_q\,\partial_r-\partial_p+
\frac{\sigma_q\,\partial_r-\sigma_r\,\partial_q}
{\sigma_q^2-\sigma_r^2}+\frac{\lambda}{2}\,\sigma_p\ \ , 
\ \ \varepsilon_{p q r}=1\ \ ,
\end{equation}     
\begin{equation}\label{31.3}
\mbox{where}\ \ \ \lambda:=\hbar\,\frac{\Lambda}{3}\ \ ,\ \ 
\partial_q:=\frac{\partial}{\partial \sigma_q}\ \ .
\end{equation}
\\
It is possible to show that the very restrictive
condition (\ref{29}) now implies that
each operator ${\cal Q}_p$ annihilates $\Psi'$
separately.
A comparison with the corresponding result for the diagonal model,
cf. eq. (2.23) of \cite{0}, reveals that,
apart from a global factor $\sqrt{\sigma_1\,\sigma_2\,\sigma_3}$,
we have an additional
factor ordering term in (\ref{31.2}), which becomes singular where two
of the $\sigma_p$ become identical\footnote{
Such a factor ordering term is well-known from earlier work, cf. \cite{csord}.}.
The occurance of the singularities
in these terms and the zeros in the measure of (\ref{sp++}),
see below, reflect the fact that the transformation (\ref{30})
between the variables $h_{p q}$ in which the quantization is performed 
and the variables $\{\sigma_p,\varphi_a \}$, which separate
the gauge degrees of freedom from the physical degrees of freedom,
is not invertible whenever two of the $\sigma_p$ take the same value.
However, this factor ordering term is just a quantum correction, which
does not affect the semiclassical limit $\hbar \to 0$, and should
be expected to appear, cf. (i) of subsection A.

To conclude this discussion, we should remark on the inner product on the
Hilbert space of wavefunctions, belonging to the two different models. 
In a first step we want to compute the auxiliary inner product (\ref{ip})
for the
non-diagonal model, which remains for 
diffeomorphism invariant wavefunctions. Let us therefore rewrite
the fundamental variabes $h_{p q}$ in the form  

\begin{equation}\label{30}
h_{p q}= \sum_{r=1}^{3}\,\Omega_{p r}\,\lambda_r\,\Omega_{q r}\ \ ,
\end{equation}
\\
where the SO(3)-matrix 
$(\Omega_{p r})$ depends on three Euler angles
$\varphi_a$. After a subsequent transformation to the $\sigma_p$
defined in (\ref{sig}) the auxiliary scalar product (\ref{ip})
transforms into

\begin{equation}\label{sp++}  
\langle \Psi | \Phi \rangle \propto
\int \mbox{d}\Omega(\varphi_a)\,
 \int \mbox{d}^3 \sigma_p
\frac{|(\sigma_1^2-\sigma_2^2)\,(\sigma_2^2-\sigma_3^2)\,
(\sigma_3^2-\sigma_1^2)|}
{{(\sigma_1\,\sigma_2\,\sigma_3)}^{3/2}}\,
   \Psi^* (\sigma_p) \, \Phi(\sigma_p)
\ \ ,
\end{equation}
\\
where $\mbox{d}\Omega(\varphi_a)$ is the invariant Haar-measure on the
group manifold of SO(3) and the variables $\sigma_p$ must be
positive and of a fixed order, e.g. $\sigma_3 >\sigma_2 >\sigma_1>0$
.\footnote{The transformation (\ref{30}) turns out to be invertible
only for a fixed order of the $\lambda_p$ (or, eqivalently, of the
$\sigma_p$), because permutations of the $\lambda_p$ may be
realized by suitable rotations with SO(3)-matrices $\Omega_{p r}$.}
The Euler angles are pure gauge
degrees of freedom and have to be gauge fixed at some arbitrary value
in the physical scalar product, which is simply done by dropping the 
factor $\int \mbox{d}\Omega$ in (\ref{sp++}), which
is just a constant prefactor, anyway. 

Remarkably, we get the same result for the effective inner product,
if we start from an auxiliary inner product in the triad variables
$e_{p a}$, which are defined via

\begin{equation}
e_{p a}\cdot e_{q a}=h_{p q}\ \ .
\end{equation}
\\
If we impose the simple inner product 

\begin{equation}\label{F1}
\langle \Psi |\Phi \rangle =\int_{\mbox{{\boldmath$R$}}^9}
\mbox{d}^9 e_{p a}\,\psi^*(e_{p a})\,\phi(e_{p a})
\end{equation}
\\
in these triad coordinates we have to eliminate the gauge freedoms
with respect to the diffeomorphism {\em and} the Gau{\ss} constraints.
To fix all these gauge freedoms, we may impose the six gauge
conditions

\begin{equation}\label{F2}
\chi^p_{\ a}=\tilde e^{p}_{\ a}=0\ \ ,\ \ p \not=a\ \ ,
\end{equation}
\\
which require the triad to be diagonal. Then the effective measure
in the scalar product (\ref{F1}) can be calculated as a Fadeev-Popov
determinant of the commutators between the generators ${\cal H}_p,
J_a'$ of the gauge transformations and the gauge conditions (\ref{F2}).
If we denote the remaining diagonal elements of $\tilde e^p_{\ a}$
by $\sigma_p$, and, moreover, take into consideration all the 
symmetries of the wavefunctions with respect to permutations
of the $\sigma_p$,
we arrive exactly at the scalar product given in (\ref{sp++}). 

In the diagonal Bianchi IX model a natural scalar product is 

\begin{equation}\label{ipd}
\langle \Psi |\Phi \rangle \propto \int_{\mbox{{\boldmath$R$}}^3}
\mbox{d}\alpha\,\mbox{d}\beta_+\,\mbox{d}\beta_-\,\Psi^*(\alpha,
\beta_{\pm})\,\Phi(\alpha,\beta_{\pm})\ \ ,
\end{equation}
\\
because the variables $\alpha, \beta_+$ and $\beta_-$ introduced in
\cite{0} are the free falling coordinates on minisuperspace for the
diagonal model. Transforming this to the $\sigma_p$-representation,
we arrive at

\begin{equation}\label{ipd+}
\langle \Psi |\Phi \rangle \propto \int_{\sigma_p >0}\,
\frac{\mbox{d}\sigma_1\,\mbox{d}\sigma_2\,\mbox{d} \sigma_3}
{\sigma_1\,\sigma_2\,\sigma_3}\,\Psi^*(\sigma_p)\,\Phi(\sigma_p)\ \ ,
\end{equation}
\\
where the three $\sigma_p$-integrals have to be performed along the 
positive real axes now. Obviously, the additional factors $
\sigma_p^2-\sigma_q^2$ occuring in (\ref{sp++}) are absent in the 
diagonal case, i.e. the measures in the scalar product of the
diagonal and the non-diagonal model differ, as was already expected
when discussing point (ii) in subsection A.

We conclude that quantization and gauge-fixing are essentially
not interchangable for the model under investigation. The 
quantization procedure followed in section II seems preferable because
it seems much nearer to
a quantization of the full, inhomogeneous field than the methods 
used for the diagonal case.
Moreover, all four constraints ${\cal H}_{\mu}$ are quantized
and in this sense treated in a similar way, as one would expect from the
viewpoint of general covariance.   

To complete our discussion of the inner product on the space of
wavefunctions,
we should give the {\em physical} inner product, which is obtained
from the {\em auxiliary} inner product (\ref{sp++}) or (\ref{ipd+}) by
eliminating
the last gauge degree of freedom corresponding to the Hamiltonian
constraint ${\cal H}_0$.

\subsection{The physical inner product on the Hilbert space of
wavefunctions}

To shorten the present paper, we want to postpone a detailed motivation
and derivation of the physical inner product to a future paper \cite{fut}.
There, we will also investigate the normalizability of the wavefunctions
constructed in section IV, and it will turn out that the physical inner
product uniquely determines one normalizable state in both of the
two cases $\Lambda = 0$ and $\Lambda >0$.
In this subsection we only want to give the final form of the physical
inner product on the space of wavefunctions and some main ideas how
to arrive there \cite{wood,FP,mar}.

The effective inner product for a constrained system 
with constraint $T=0$ is generally 
obtained by firstly choosing a suitable gauge condition $\chi$, and
secondly calculating the following expectation value with respect
to the auxiliary (or `kinematic') inner product: 

\begin{equation}\label{IP1}
\langle \Psi || \Phi \rangle_{phys}=
\langle \Psi |\,|J|\,\delta(\chi) |\Phi \rangle_{aux}\ \ \ \ ,\ \ \ \
\mbox{with}\ \ \ \ J:=i\, [T,\chi]\ \ .
\end{equation}
\\
Here $J$ is the so-called Faddev-Popov
determinant. This definition makes sense whenever $T$ is a first order
derivative operator on configuration space, because then $|J|$ will be
a positive real number and (\ref{IP1}) defines a positive hermitian
product. However, in the present case $T=\hat {\cal H}_0$ is a second
order derivative operator, and we have to deal with two problems:

\begin{itemize}

\item [(i)] At first sight, $|J|$ seems to be a mysterious quantity,
because $J$ itself is a first order derivative operator on configuration
space. However, since $J$ is a self-adjoint operator, we may decompose
any state $\Psi$ in terms of eigenstates of $J$ to arrive at a very 
natural notion of $|J|$. This operator $|J|$ then is itself self-adjoint,
and, moreover, {\em positive}. 

\item [(ii)] In general, $|J|$ and $\delta(\chi)$ will not commute,
so $\langle \Psi || \Phi \rangle_{phys}$ will neither be hermitian, nor positive.
This would give a rather miserable physical inner product. Surprisingly,
there exists a small set of gauge conditons $\chi$ such that the
gauge condition commutes with the Faddev-Popov determinant, i.e.
such that $\chi$ solves

\begin{equation}\label{IP2}
 \big [ [T,\chi],\chi \big ]=0\ \ .
\end{equation}
\\
It seems that this set of gauge conditions does not only exist
for the case under study, $T=\hat {\cal H}_0$, but for any physically
interesting case of constraints which are described by second order
derivative operators. Using these special gauge conditions the physical
inner product (\ref{IP1}) obviously {\em is} poitive and hermitian.

\end{itemize}

Having solved these two problems, we are now ready to perform the last step
in the construction of the physical inner product
for the non-diagonal Bianchi IX
model. It is easily seen that the gauge condition

\begin{equation}\label{IP3}
\chi(\sigma_1,\sigma_2,\sigma_3)=
\frac{2}{\hbar}\,\mbox{ln} \frac{\sigma_3}{\sigma_{3,0}} 
\end{equation}
\\
is a solution to eq. (\ref{IP2}). It fixes the variable
$\sigma_3$ to have a specific value $\sigma_3 \equiv \sigma_{3,0}$.
Performing suitable coordinate transformations the associated inner
product finally takes the following form:

\begin{equation}\label{IP4}
\langle \Psi || \Phi \rangle_{\sigma_{3}}=
\int\limits_{0}^{+\infty}\,\mbox{d}\beta_- \int\limits_{0}^{\infty}
\,\mbox{d}v\ \bar \Psi^*\,\big |i\,\hbar\,\frac{\partial}{\partial v} 
\big |\,
\bar \Phi\ \ \ \ ,\ \ \ \ \mbox{with}
\end{equation}
\begin{equation}\label{IP5}
\bar \Psi=\left [
\frac{(\sigma_1^2-\sigma_2^2)\,(\sigma_2^2-\sigma_3^2)\,
(\sigma_3^2-\sigma_1^2)}{\sigma_1\,\sigma_2\,\sigma_3} \right ]^{1/2}\ \Psi
\ \ \ \ ,\ \ \ \ v=a^3=\left ( \frac{\hbar}{2} \right )^{3/2}\!\!\!
\sqrt{\sigma_1\,\sigma_2\,\sigma_3}\ \ \ \ ,\ \ \ \ \beta_-=
\frac{1}{2 \sqrt{3}}\ \mbox{ln} \frac{\sigma_2}{\sigma_1}\ \ .
\end{equation}
\\
This inner product has now an intrinsic
`time'-parameter $\sigma_{3}$, which must
be given to compute probability distributions with respect to $\beta_-$
and $v$, the remaining physical degrees of freedom.

\section{Solutions generated from the Chern-Simons functional}

\subsection{Ashtekar representation of the Chern-Simons state}

As pointed out in section II, we can find a special
exact quantum state
for the non-diagonal Bianchi IX model with a non vanishing cosmological
constant, if we are able to determine a function $\psi'$ lying in
the kernel of ${\cal Q}^p_{\ a}$ defined in (\ref{19}). To proceed in 
this way, let us consider the Fourier transformed version of 
(\ref{29}), where it is essential to perform not a customary,
but a generalized
Fourier transform introduced in \cite{0}:

\begin{equation}\label{32}
\psi'(\tilde e^p_{\ a})\,\propto\,
\int_{\Sigma^9}\,\mbox{d}^9 \!A_{p a}\,\exp\left
 [\frac{2}{\hbar}\,A_{p a}\,
\tilde e^p_{\ a} \right ]\ \tilde \Psi(A_{p a})\ \ .
\end{equation}
\\
Here $\Sigma^9\!\subset\,${\bf{C}}$^9$ is any 9 dimensional manifold
in the 18-dimensional space of the complex $A_{p a}$,
which allows for partial
integration without getting any boundary terms. Then
$\psi'(\tilde e^p_{\ a})$ is a solution to (\ref{29}), if
$\tilde \Psi(A_{p a})$ solves the following set of equations:

\begin{equation}\label{33}
\left [
m^{p q}\,A_{q a}-\frac{1}{2}\,\varepsilon^{p q r}\,
\varepsilon_{a b c}\,A_{q b}\,A_{r c}+\frac{\lambda}{2}\,
\frac{\partial}{\partial A_{p a}}
\right ]\,
\tilde \Psi(A_{p a})=0\ \ .
\end{equation}
\\
A special solution to (\ref{33}) is the Bianchi IX restriction of the
well known Chern-Simons functional:

\begin{equation}\label{34}
\tilde \Psi_{\mbox{{\tiny $C\!S$}}}=
\exp \left [
\frac{1}{\lambda}\,\big (
-A_{p a}\,m^{p q}\,A_{q a}+2\,\det(A_{p a}) \big )\right ]\ \ .
\end{equation}
\\
In the following we shall be interested in the
transformation of this state to the metric representation,
i.e. in the evaluation of integrals of the form (\ref{32}). 
Topologically different choices of $\Sigma^9$ (i.e. choices which
cannot be deformed into each other without passing a singularity)
will lead to different states in the metric representation.

\subsection{Metric representation of the Chern-Simons state}

It will be convenient to introduce new variables 

\begin{equation}\label{35}
\kappa^p_{\ a}:=\frac{\lambda}{\hbar}\,\tilde e^p_{\ a}=
\frac{\Lambda}{3}\,\tilde e^p_{\ a}\ \ .
\end{equation}
\\
Then the wavefunction in the metric representation takes the form

\begin{equation}\label{36}
\Psi'_{\mbox{{\tiny $C\!S$}}}\,\propto\,
\int_{\Sigma^9}\,\mbox{d}^9 \! A_{p a}\,\exp \left [
\frac{2}{\lambda} \bigg (
\kappa^p_{\ a}\,A_{p a}-\frac{1}{2}\,A_{p a}\,m^{p q}\,A_{q a}
+\,\det(A_{p a})
\bigg ) \right ]\ \ ,
\end{equation}
\\
where suitable integration manifolds $\Sigma^9$ remain to be
determined. Here and in the following, constant prefactors, and
also prefactors depending on $\lambda$, are absorbed in the 
proportionality sign `$\propto$', since such a factor will always
remain as a freedom of the wavefunction.
Surprisingly, six of the nine $A_{pa}$-integrals turn out to be
Gaussian integrals and can be evaluated analytically,
namely the
integrals with respect to $A_{1 1},A_{1 2},A_{1 3},A_{2 1},
A_{2 2}$ and $A_{2 3}$. Lengthy calculations finally give the
result

\begin{equation}\label{37}
\Psi'_{\mbox{{\tiny $C\!S$}}}\,\propto\,
\int_{\Sigma^3}\,\frac{ \mbox{d}x\, \mbox{d}y\, \mbox{d}z}{1-{\vec r\,}^2} 
\exp \left [ \frac{1}{\lambda} \,\bigg (\!
-{\vec r\,}^2+2\,{\vec \kappa}_3\!\cdot\!{\vec r}+\frac{
{\vec \kappa}_1^2+{\vec \kappa}_2^2+
2\, {\vec \kappa}_1\!\!\times\!{\vec \kappa}_2\!\cdot\!{\vec r}-
({\vec r}\!\cdot\!{\vec \kappa}_1)^2-
({\vec r}\!\cdot\!{\vec \kappa_2})^2}{1-{\vec r\,}^2} \bigg ) \right ]\ \ ,
\end{equation}
\\
where we have introduced the abbreviations 
${\vec r}=(x,y,z):=(A_{3 1},A_{3 2},A_{3 3})$ and
${\vec \kappa}_p:=(\kappa^p_{\ 1},\kappa^p_{\ 2},\kappa^p_{\ 3})$.
It is clear that the 3-direction is distinguished by the order in which
the integrals are performed.
As we know from section II, $\Psi'_{\mbox{{\tiny $C\!S$}}}$
according to (\ref{37}) is automatically Gau{\ss}- and diffeomorphism
invariant, because it is a solution to (\ref{29}). This means that
$\Psi'_{\mbox{{\tiny $C\!S$}}}$ can only depend on the eigenvalues
$\lambda_p$ of the three-metric $h_{p q}$, or, equivalently, on the
$\sigma_p$ introduced in (\ref{sig}). So we are free to choose a
diagonal gauge for the triad in (\ref{37}), i.e. we may take

\begin{equation}\label{dg}
(\kappa^p_{\ a})=\mbox{diag}\,(\kappa_1,\kappa_2,\kappa_3)\ \ \ \ ,
\ \ \mbox{with}\ \ \ \ \kappa_3 \geq \kappa_2 \geq \kappa_1\ \ ,
\end{equation}
\begin{equation}\label{38}
\mbox{where}\ \ \ \ 
\kappa_p:=\frac{\lambda}{2}\,\sigma_p=\frac{\Lambda}{12}\,\sqrt{
\lambda_q\,\lambda_r}\ \ ,\ \ \varepsilon_{p q r}=1\ \ .
\end{equation}
\\
This special ordering of the three $\kappa_p$ (or, equivalently,
of the $\lambda_p$) is always possible, because  
permutations of the diagonal elements of a diagonal metric
$h_{p q}$ correspond to diffeomorphisms, which leave the wavefunctions
invariant. Moreover, such an ordering is even necessary to have
a unique mapping $(\kappa^p_{\ a}) \rightarrow (\kappa_1,\kappa_2,
\kappa_3)$ under Gau{\ss}- and diffeomorphism transformations,
and we will adopt the convention (\ref{dg}) 
throughout the following. In the gauge (\ref{dg}) the integral
(\ref{37}) takes the form 

\begin{equation}\label{39}
\Psi'_{\mbox{{\tiny $C\!S$}}}\,\propto\,
\int_{\Sigma^3}\,\frac{ \mbox{d}x\, \mbox{d}y\, \mbox{d}z}{1-r^2} 
\exp \left [ \frac{1}{\lambda} \,\bigg (\!
-r^2+2\, \kappa_3\, z+\frac{
 \kappa_1^2+ \kappa_2^2+
2\, \kappa_1 \kappa_2 \,z-
\kappa_1^2\,x^2-\kappa_2^2\,y^2
}{1-r^2} \bigg ) \right ]\ \ , 
\end{equation}
\\  
where $r^2:=x^2+y^2+z^2$. We will see in sections V and VI that there
are choices for $\Sigma^3$ in (\ref{39}) leading to asymmetric
wavefunctions under (formal) permutations of the 
$\kappa_p$.\footnote{
In complete analogy there are eigenstates of the effective
Hamiltonian (\ref{H8}) of a 2-dimensional harmonic oscillator with
$L=0$, which do {\em not} obey $\Psi(q)=\Psi(-q)$, but these are of
no physical interest.} However, as we know from section III
(cf. after eq. (\ref{D6}) with $\kappa_p = \frac{\lambda}{2}\,\sigma_p$),
only
wavefunctions which are symmetric under permutations of the $\kappa_p$
and under reflexions $\kappa_p \rightarrow -\kappa_p,
\, \kappa_q \rightarrow -\kappa_q,\,
\kappa_r \rightarrow \kappa_r$ are
of interest to us, and this will be an important restriction
to select the physically interesting wavefunctions.

\section{Asymptotic forms of the Chern-Simons integral}

In order to get information about possible integration manifolds
$\Sigma^3$ which can be used in eq. (\ref{39}) we begin by discussing
the asymptotic behavior of the Chern-Simons integral in several asymptotic
regimes. It will thereby become obvious that we deal with five
linearly independent solutions,
and the integration contours will be given in section VI afterwards.

\subsection{The semiclassical limit $\hbar \to 0$}

Let us first of all examine the semiclassical behavior of the solutions
described by (\ref{39}). Surely, the semiclassical limit could have also
been discussed by starting from the nine dimensional integral (\ref{36}),
but, of course, an expansion of the three dimensional integral (\ref{39})
is much simpler. The saddle point form of (\ref{39}) is displayed nicely
by writing it in the form

\begin{equation}\label{40}
\Psi'_{\mbox{{\tiny $C\!S$}}}\,\propto\,
\int\,\frac{ \mbox{d}x\, \mbox{d}y\, \mbox{d}z}{1-r^2}\, 
\exp \left [\frac{3\,F}{\hbar\,\Lambda} \right ]\ \ ,
\end{equation}
\begin{equation}\label{40+}
\mbox{with}\ \ 
F:=-r^2+2\, \kappa_3\, z+\frac{\kappa_1^2+\kappa_2^2+2\,\kappa_1 \kappa_2\, z
-\kappa_1^2\, x^2-\kappa_2^2\, y^2}{1-r^2}\ \ ,
\end{equation}
where we have inserted $\lambda$ according to (\ref{31.3}). Since the
$\kappa_p$ defined in (\ref{38}) do not depend on $\hbar$, we approach 
a Gaussian integral in the limit $\hbar \to 0$, $\Lambda$ fixed,
which has to be
evaluated at one of the saddle points of the exponent. However, {\em which}
of the possible saddle point contributions arises for the integral under
consideration is determined by the integration surface $\Sigma^3$ and
requires a detailed discussion of the contours of steepest descent.
Here we only want to give the {\em possible}
asymptotic results for $\hbar \to
0$ which may be realized by suitable choices of the integration contours.
The integration contours are discussed in section VI.

The saddle points of the exponent are obtained by solving the equations  

\begin{equation}\label{41}
\frac{\partial F}{\partial x}=0\ \ ,\ \ 
\frac{\partial F}{\partial y}=0\ \ ,\ \ 
\frac{\partial F}{\partial z}=0\ \ . 
\end{equation}
\\
One can show that for $\sigma_1,\sigma_2,\sigma_3$ pairwise different,
i.e. in particular on the sector $\sigma_3 > \sigma_2 >\sigma_1 >0$
of interest,
the only solution to (\ref{41}) is given by

\begin{equation}\label{42}
x=y=0\ \ ,\ \ 
(z-\kappa_3)\,(1-z^2)^2=(\kappa_1+\kappa_2 z)\,(\kappa_2+\kappa_1 z)\ \ .
\end{equation}
\\
A comparison with the diagonal model \cite{0} reveals that for $x=y=0$
the exponents of (\ref{40}) and of the one dimensional integral (4.17)
of \cite{0} become identical. Consequently, also the saddle point
equations with respect to $z$ coincide, cf. eqs. (\ref{42}) and (4.7)
of \cite{0}. As a result, the semiclassical actions of the wavefunctions
approached in the limit $\hbar \to 0$ must be the same, since they are
given as the saddle point values of the integrand's exponents. This is
of course exactly what we expected to happen, because
the Wheeler-DeWitt equations for the classically equivalent diagonal and 
non-diagonal Bianchi IX model
give rise to the same Hamilton Jacobi
equation in the limit $\hbar \to 0$. However, the quantum corrections
to this leading order behavior are rather different, firstly because
of the different dimensionality of the integrals, and secondly because
of a new prefactor to the exponential function. Performing a Gaussian
saddle point approximation, we get for the non-diagonal case

\begin{equation}\label{43}
\Psi'_{\mbox{{\tiny $C\!S$}}}
\, \propto^{^{\!\!\!\!\!\!\!\!\!\!\!
\mbox{{\scriptsize $\hbar\! \to \!0$}}}}
\,
\left [
\frac{(-2 \pi \lambda)^3}{\det \Bigl ( 
\mbox{\large{$\frac{\partial^2 F}{\partial r_a\,
\partial r_b}$}} \Bigr )} \right ]^{\frac{1}{2}}\ 
\frac{e^{F/ \lambda}}{1-r^2}\ \bigg |_{\vec r={\vec r}_{s}}
\propto
\,\left [\,
\frac{\partial^2 F}{\partial x^2}\,
\frac{\partial^2 F}{\partial y^2}\,
\frac{\partial^2 F}{\partial z^2}\, \right ]^{-\frac{1}{2}}\ 
\frac{e^{F/ \lambda}}{1-r^2}\ \bigg |_{\vec r={\vec r}_{s}}\ \ .
\end{equation}
\\
Here the second proportionality stems from a special property of
the exponent $F$, namely the fact that the mixed derivatives of $F$
vanish for $x=y=0$. Comparing this with the asymptotic results obtained
in the diagonal case \cite{0}, we find that corresponding results differ
only by a factor

\begin{equation}\label{44}
\gamma:=\left [(1-z^2)\,
\frac{\partial^2 F}{\partial x^2}\,
\frac{\partial^2 F}{\partial y^2}\right ]_{x=y=0,\,z=z_s}^{
-\frac{1}{2}}\ \ .
\end{equation}
\\
Since we already know the asymptotic results for the diagonal Bianchi
IX model, we only have to expand this additional factor at
the saddle points to get
the quantum corrections for the non-diagonal Bianchi IX model. As in
the diagonal case, we therefore have to deal with a saddle point equation
of fifth order, so analytical results are only available in additional
asymptotic regimes. Due to the existence of five saddle points one has
five linearly independent solutions.

\subsubsection{The limit $\Lambda \to 0$}

One asymptotic regime which allows for an expansion of the five
saddle points is that of small cosmological constant
$\Lambda \to 0$. Since we are in the
additional limit $\hbar \to 0$, the quantity $\lambda=\frac{\hbar \Lambda}
{3}$ defined in (\ref{31.3}) tends to zero, too. 
In this limit one saddle point is given by
 
\begin{equation}\label{45}
z\ \ \sim^{^{\!\!\!\!\!\!\!\!\!\!\!
\mbox{{\scriptsize $\lambda\! \to \!0$}}}}
\,\frac{1}{2}\,\sigma_3\, \lambda
\ \, \to^{^{\!\!\!\!\!\!\!\!\!\!\!
\mbox{{\scriptsize $\lambda\! \to \!0$}}}} 0\ \ .
\end{equation}
\\
The contribution from this saddle point yields

\begin{equation}\label{46}
\Psi'_{\mbox{{\tiny $C\!S$}}}
\  \propto^{^{\!\!\!\!\!\!\!\!\!\!\!
\mbox{{\scriptsize $\lambda\! \to \!0$}}}}
\,const.\ \ ,
\end{equation}
\\
and therefore approaches the well-known wormhole state.\footnote{
We remind the reader that the prime at $\Psi'_{\mbox{{\tiny $C\!S$}}}$
denotes the fact that this wavefunction has to be mutliplied with
$\Psi_{\mbox{{\tiny $W\!H$}}}$ defined in (\ref{18}) to become a solution
of the Wheeler DeWitt equation in the metric representation.}

Four further saddle points asymptotically lie at\footnote{
The following expansions are only valid for $\sigma_2>\sigma_1$, but
we restricted ourselves to this case anyway.}

\begin{equation}\label{47}
z\ \ \sim^{^{\!\!\!\!\!\!\!\!\!\!\!
\mbox{{\scriptsize $\lambda\! \to \!0$}}}}
\,-1 \pm \frac{\lambda}{4}\,(\sigma_2-\sigma_1)\ \ ,\ \ 
z\ \ \sim^{^{\!\!\!\!\!\!\!\!\!\!\!
\mbox{{\scriptsize $\lambda\! \to \!0$}}}}
\,1 \pm \frac{\lambda}{4}\,(\sigma_2+\sigma_1)\ \ ,\ \ 
\end{equation}
\\
giving rise to an asymptotic behavior

\begin{equation}\label{48}
\Psi'_{\mbox{{\tiny $C\!S$}}}
\  \propto^{^{\!\!\!\!\!\!\!\!\!\!\!
\mbox{{\scriptsize $\lambda\! \to \!0$}}}}
\,\frac{e^{-\sigma_3 \mp (\sigma_2-\sigma_1)}}{\sqrt{
(\sigma_2-\sigma_1)(\sigma_3 \pm \sigma_2)(\sigma_3 \mp \sigma_1)}}\ \ ,
\ \
\Psi'_{\mbox{{\tiny $C\!S$}}}
\  \propto^{^{\!\!\!\!\!\!\!\!\!\!\!
\mbox{{\scriptsize $\lambda\! \to \!0$}}}}
\,\frac{e^{\sigma_3 \mp (\sigma_2+\sigma_1)}}{\sqrt{
(\sigma_1+\sigma_2)(\sigma_3 \mp \sigma_1)(\sigma_3 \mp \sigma_2)}}\ \ ,
\end{equation}
\\
respectively. These solutions are the generalizations of the
diagonal analogues (2.25) and (2.26) given in \cite{0}. The singularities
which occur in the denominators of the asymmetric solutions of
(\ref{48}) are
cancelled by the weight function of the inner product (\ref{sp++})
and therefore do not constitute any physical problems.

\subsubsection{The limit $\kappa \to \infty$}

A second regime which allows for an analytical expansion of the semiclassical
limit $\hbar \to 0$ is the case $\kappa \to \infty$, where the `mean'
$\kappa$ is understood as $\kappa:=(\kappa_1\,\kappa_2\,\kappa_3)^{1/3}$.
The three $\kappa_p$ where defined in eq.~(\ref{38}).
For a fixed cosmological constant, this is the limit of large
overall scale paramter. In this limit,
two of the saddle points behave according to

\begin{equation}\label{50}
z\ \ \sim^{^{\!\!\!\!\!\!\!\!\!\!\!
\mbox{{\scriptsize $\kappa\! \to \!\infty$}}}}
\ \pm \,\sqrt{-\frac{\kappa_1\,\kappa_2}{\kappa_3}}\ \ ,
\end{equation}
\\
and turn out to be complex-valued. This gives rise
to a complex action of the wavefunction, describing a Universe with
a Lorentzian signature of the 4-metric, cf. \cite{0}. The leading order
behavior for these solutions is given by

\begin{equation}\label{51}
\Psi'_{\mbox{{\tiny $C\!S$}}}
\ \ \propto^{^{\!\!\!\!\!\!\!\!\!\!\!
\mbox{{\scriptsize $\kappa\! \to \!\infty$}}}}
\, \kappa^{-\frac{9}{4}}\,\exp \left [
\pm \frac{4 i \,\kappa^{3/2}}{\lambda} \right ]\ \ ;
\end{equation}
\\
for a more detailed asymptotic result we refer to \cite{0}. Three further
saddle points have an asymptotic expansion 

\begin{equation}\label{52}
z\ \ \sim^{^{\!\!\!\!\!\!\!\!\!\!\!
\mbox{{\scriptsize $\kappa\! \to \!\infty$}}}}
 -\frac{\kappa_1}{\kappa_2}\ \ ,\ \
 z\ \ \sim^{^{\!\!\!\!\!\!\!\!\!\!\!
\mbox{{\scriptsize $\kappa\! \to \!\infty$}}}}
 -\frac{\kappa_2}{\kappa_1}\ \ ,\ \
z\ \ \sim^{^{\!\!\!\!\!\!\!\!\!\!\!
\mbox{{\scriptsize $\kappa\! \to \!\infty$}}}}
 1+\kappa_3 \ \ ,
\end{equation}
\\
and lead to wavefunctions of the form

\begin{displaymath}
\Psi'_{\mbox{{\tiny $C\!S$}}}
\  \propto^{^{\!\!\!\!\!\!\!\!\!\!\!
\mbox{{\scriptsize $\kappa\! \to \!\infty$}}}}
\,\ \bigl [ \left  (\kappa_2^2-\kappa_1^2 \right )\,
\left (\kappa_3^2-\kappa_2^2 \right ) \bigr ]^{-\frac{1}{2}}\ 
e^{\kappa_2^2/\lambda}\ \ ,\ \ 
\Psi'_{\mbox{{\tiny $C\!S$}}}
\  \propto^{^{\!\!\!\!\!\!\!\!\!\!\!
\mbox{{\scriptsize $\kappa\! \to \!\infty$}}}}
\,\ \bigl [ \left  (\kappa_2^2-\kappa_1^2 \right )\,
\left (\kappa_3^2-\kappa_1^2 \right ) \bigr ]^{-\frac{1}{2}}\ 
e^{\kappa_1^2/\lambda}\ \ ,
\end{displaymath}
\begin{equation}\label{53} 
\Psi'_{\mbox{{\tiny $C\!S$}}}
\  \propto^{^{\!\!\!\!\!\!\!\!\!\!\!
\mbox{{\scriptsize $\kappa\! \to \!\infty$}}}}
\,\ \bigl [ \left  (\kappa_3^2-\kappa_1^2 \right )\,
\left (\kappa_3^2-\kappa_2^2 \right ) \bigr ]^{-\frac{1}{2}}\ 
e^{\kappa_3^2/\lambda}\ \ ,\ \ 
\end{equation}
\\
respectively. These saddle point contributions are known from the
diagonal model for the `asymmetric' solutions introduced there;
however, here we have additional prefactors, which become divergent
on the symmetry lines $\sigma_p =\sigma_q,\,p \not= q$, but again these
singular terms are cancelled by the measure of the scalar product (\ref{sp++}).

\subsection{Solutions to the vacuum model approached in the limit 
$\Lambda \to 0$}

\begin{figure}
\begin{center}
\leavevmode
\hskip 0 cm
\epsfxsize 10cm
\epsfbox{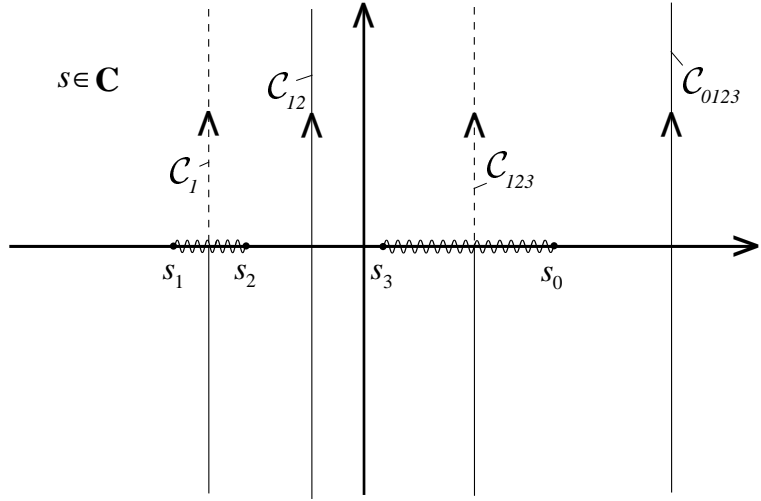}
\end{center}
\caption{Integration curves in the complex $s$-plane which lead to linearly
independent vacuum solutions. A dashed line indicates that one has
to evaluate the integrand in the second Riemannian branch after
having crossed one of the cuts. The two cuts are represented by
wavy lines.}
\label{fig1}
\end{figure}

In order to get solutions of the non-diagonalnon-diagonal Bianchi IX model for
$\Lambda  \to 0$ (without taking the semicalssical limit $\hbar \to 0$
first) 
we now want to discuss the behavior of the Chern-Simons
integral in this limit. While one of the solutions approached for 
$\Lambda \to 0$ again turns out to be the wormhole state (\ref{18}),
the other four solutions are {\em not} given by the asymptotic results
(\ref{48}), as one might think at the first sight. The reason for this
is the fact that we are not allowed to perform the usual Gaussian 
saddle point expansion, since the exponent, and in
particular the prefactor to the exponential function, become singular 
in the limit $\Lambda \to 0$, where $r$ approaches $\pm 1$. Comparing
the situation with the diagonal model, where a saddle point expansion
in the same limit actually {\em was} allowed,
one might ask for the difference
between the two models that forbids for an analogous expansion in the present
case. The answer to this question lies hidden in the prefactor to the
exponential function: while we dealed with an integrable square-root
singularity
in the diagonal case, we here have to integrate into a singularity
of first order, even in the case $\Lambda =0$, and this requires 
much caution. In fact, it turns out
that the calculation of the limit $\Lambda  \to 0$ of the Chern-Simons
integral is very subtle and worth to be discussed in
a separate section in appendix A. Moreover, we show in appendix B that
the solutions derived in appendix A can be written in another, very nice
and compact form, which reads

\begin{equation}\label{52+}
\Psi'_{\varrho}
\ \, \sim^{^{\!\!\!\!\!\!\!\!\!\!\!
\mbox{{\scriptsize $\Lambda\! \to \!0$}}}}
\,
=\int_{\mbox{\small{${\cal C}_{\varrho}$}}}
\mbox{d}s\,e^s \,
\prod_{\nu=0}^{3}\,(s-s_{\nu})^{-\frac{1}{2}}\ \ .
\end{equation}
\\
Here the integration contour ${\cal C}_{\varrho}$ is one of the
four curves shown in figure \ref{fig1}, and the $s_{\nu}$ are special
sums of the $\sigma_p$, defined by

\begin{equation}\label{sss}
s_0=\sigma_1+\sigma_2+\sigma_3\ \ ,\ \
s_1=\sigma_1-\sigma_2-\sigma_3\ \ ,\ \
s_2=-\sigma_1+\sigma_2-\sigma_3\ \ ,\ \
s_3=-\sigma_1-\sigma_2+\sigma_3\ \ .
\end{equation}
\\
For a detailed discussion of these wavefunctions
and a comment on their symmetries with respect to permutations
of the $\sigma_p$ we refer to appendix A and B. 

Altogether, we find in the limit $\Lambda \to 0$ 
again five linearly independent solutions associated with
the Chern-Simons wavefunction (\ref{36}).

\vspace{0.5 cm}

\hspace{2 cm}

\section{Three dimensional integration manifolds for five exact 
solutions}

In this section we want to define explicitely five integration contours
$\Sigma^3$ for the integral (\ref{39}), leading to five linearly
independent solutions of the Wheeler DeWitt equation (\ref{31.1}).
To have a better view over
the symmetries of the solutions, let us, for a moment, go back to the
nine dimensional integral representation (\ref{36}). It can be verified
easily that the two different integration surfaces

\begin{equation}\label{54}
\Sigma^9_{\pm}:=\bigl \{ (A_{p a})\,\epsilon\,\mbox{\bf{C}}^9\,|
A_{p a}\,\epsilon\,\mbox{\bf{R}}\,e^{\pm i \pi/6} \bigr \}
\end{equation}
\\
lead to a finite integral, because the cubic terms in the exponent
of (\ref{36}) remain purely imaginary, while all the quadratic
terms have a negative
real part. Moreover, since the only border of this surfaces lies at infinity,
it is clear that the integral (\ref{36}),
performed over $\Sigma^9_{\pm}$, must
lead to a solution of the non-diagonal Bianchi IX model.
In particular, we then know
that these wavefunctions are Gau{\ss}- and diffeomorphism
invariant, and we may again choose the diagonal gauge (\ref{dg})
for the matrix $(\kappa^{p}_{\ a})$. Then (\ref{36}) takes
the form

\begin{equation}\label{55}
\Psi'_{\mbox{{\tiny $C\!S$}},\pm}\,\propto\,
\int_{\Sigma^9_{\pm}}\,\mbox{d}^9 \! A_{p a}\,\exp \left [
\frac{2}{\lambda} \bigg (
\kappa_1\,A_{1 1}+\kappa_2\,A_{2 2}+\kappa_3\,A_{3 3}
-\frac{1}{2}\,A_{p a}\,m^{p q}\,A_{q a}
+\,\det(A_{p a})
\bigg ) \right ]\ \ .
\end{equation}
\\
If we now consider a formal
permutation $\kappa_1 \leftrightarrow \kappa_2$ in
(\ref{55}), we can re-establish the original integral by a
suitable
transformation in the $A_{p a}$-space, namely $A_{1 1}
\leftrightarrow A_{2 2},\,A_{3 2}\leftrightarrow A_{1 3},\,A_{2 3}
\leftrightarrow A_{3 1}$. Since all integration variables $A_{p a}$
are integrated along the same axes in the complex $A_{p a}$-planes, the
integration contour remains unchanged under this coordinate transformation
and we regain the integral (\ref{55}). Furthermore, 
the two integrals (\ref{55}) are invariant under a substitution
$\kappa_1 \to -\kappa_1,\,\kappa_2 \to -\kappa_2$,which may be seen by
a transformation $A_{1 a} \to - A_{1 a},\,A_{2 a} \to -A_{2 a}$ in the
$A_{p a}$-space. Thus we conclude that both wavefunctions are completely
symmetric under arbitrary permuations of the $\kappa_p$, and also
under reflexions $\kappa_p \to -\kappa_p,\,
\kappa_q \to -\kappa_q,\,\kappa_r \to \kappa_r$, a symmetry property
which is posessed only by two of the five solutions which are
generated by the Chern-Simons functional.

Performing the six Gaussian integrations which lead from (\ref{36})
to (\ref{39}), we reach the following three dimensional integration
surface for the integral (\ref{39}):

\begin{equation}\label{56}
\Sigma^3_{\pm}:=\bigl \{ \vec r\,\epsilon\,\mbox{\bf{C}}^3\,|
r_a\,\epsilon\,\mbox{\bf{R}}\,e^{\pm i \pi/6} \bigr \}\ \ .
\end{equation}
\\
In the following it will be convenient to introduce the two superpositions

\begin{equation}\label{57}
\Psi'_{\mbox{{\tiny $W\!H$}},\Lambda}
\propto\Psi'_{\mbox{{\tiny $C\!S$}},+}+
\Psi'_{\mbox{{\tiny $C\!S$}},-}\ \ ,\ \ 
\Psi'_{0123,\Lambda}\propto\Psi'_{\mbox{{\tiny $C\!S$}},+}-
\Psi'_{\mbox{{\tiny $C\!S$}},-}\ \ ,\ \ 
\end{equation}
\\
and also new integration variables

\begin{equation}\label{sc}
x=r\,\sqrt{1-\eta^2}\,\cos \varphi\ \ ,\ \ 
y=r\,\sqrt{1-\eta^2}\,\sin \varphi\ \ ,\ \ 
z=r\,\eta\ \ ,
\end{equation}
\\
for the $x,y,z$-integral, which may be understood as spherical coordinates
with $\eta=\cos \vartheta$. For the transformation (\ref{sc}) the
volume element simply transforms like $\mbox{d}x\, \mbox{d}y\, \mbox{d}z=
r^2\mbox{d}r\, \mbox{d}\eta\, \mbox{d}\varphi$, and the condition 
$r^2=x^2+y^2+z^2$ is obeyed even for complex values of $r,\eta,\varphi$.
With this new parameterization, the integration surfaces for the two
solutions defined in (\ref{57}) can now be written in the following
form:

\begin{equation}\label{59.1}
\Sigma^3_{\mbox{{\tiny $W\!H$}}}
=\bigl \{ \vec r\,\epsilon\,\mbox{\bf{C}}^3\,|
-1<\eta < 1,\,0<\varphi<2 \pi,\,r\,\epsilon\,
\mbox{\bf{R}}+i\, \varepsilon \bigr \}\ \ , \ \ \varepsilon>0\ \ ,
\end{equation}
\begin{equation}\label{59.2}
\Sigma^3_{0123}
=\bigl \{ \vec r\,\epsilon\,\mbox{\bf{C}}^3\,|
-1<\eta < 1,\,0<\varphi<2 \pi,\,r\,\epsilon\,
K(1,R) \bigr \}\ \ , \ \ R<2\ \ .
\end{equation}
\\
In (\ref{59.2}) $K(1,R)$ denotes a pole integral around $r=1$ with
radius $R$, cf. appendix A. Let us try to interpret these
integration surfaces: in the first case (\ref{59.1}) we simply integrate
over a sphere with radius $r$ in the complex $\vec r$-space;
however, this
radius is not chosen real-valued as usual,
but slightly imaginary to avoid the
singularities of the integrand lying at $r=\pm 1$. The integration
manifold $\Sigma^3_{0123}$ also describes an
integral over a full sphere, but
now the radius describes itself a circle round the singularity at $r=+1$
in the complex $r$-plane.
While the two parameterizations given in (\ref{59.1}) and (\ref{59.2})
will prove useful for an asymptotic expansion in the limit
$\Lambda \to 0$, which is performed in appendix A, they now  
allow for a further study of the semiclassical behavior of the corresponding
solutions.

If we consider the wavefunction $\Psi'_{0123,\Lambda}$
in the semiclassical
limit $\hbar \to 0$, we get the dominant contributions to the integral
for $\eta=\pm 1$, so that the integrand becomes independent of $\varphi$.
The remaining $r$-variable then coincides with the $\pm z$-variable,
which was the only integration variable in the diagonal Bianchi IX
model discussed in \cite{0}. For a discussion of the curves of
steepest descent in the $z$-plane we refer to this earlier work. However,
if one has finally deformed a desired integration contour into the
curves of steepest descent, there remains one remarkable difference
between the two models: while in the diagonal case we had a cut in the
complex $z$-plane for $|\mbox{Re}\, z|>1$, this cut is absent in the present
non-diagonal case, owing to the different prefactors to the exponential
function. So integrals along parts of the real axis of the
$z$-plane, which cancelled in the diagonal case, may now be different
from zero, and vice versa. A detailed discussion of the curves of steepest
descent and the resulting asymptotic contributions finally gives the
following result for the wavefunctions $\Psi$, being connected with
$\Psi'$ via (\ref{26}):

In the semiclassical limit $\hbar \to 0$ the wavefunction
$\Psi_{0123,\Lambda}$ shows a no-boundary behavior for $a \to 0$,
i.e. the 4-manifolds which correspond to the semiclassical
trajectories in minisuperspace are regular at the point $a=0$. For
$a \to \infty$, the integration surface $\Sigma^3_{0123}$ picks
up several saddle point contributions: the two complex conjugate
saddle points (\ref{50}) contribute as well as the two real-valued,
negative saddle points in (\ref{52}). These latter two saddle points
are the reason why $\Psi_{0123,\Lambda}$ is {\em not} normalizable
in the physical inner product (\ref{IP4}), as will be shown
in a future paper \cite{fut}. 

The solution $\Psi_{\mbox{\tiny{$W \! H$}},\Lambda}$ is a generalization
of the wormhole state (\ref{17}) for a positive
cosmological constant $\Lambda >0$. For $a \to \infty$, we again
gather contributions from the two complex conjugate saddle points
(\ref{50}), and now, in addition, from the real-valued
{\em positive} saddle
point $z \sim 1+\kappa_3$ of (\ref{52}). In \cite{fut} we will see
that these latter three saddle points give rise to
finite contributions 
in the physical norm of the wavefunction. 
We will show this by evaluating the physical norm
according to (\ref{IP4}) in the semiclassical
limit $\hbar \to 0$. As a result,
$\Psi_{\mbox{\tiny{$W \! H$}},\Lambda}$ turns out to be the
only Chern-Simons-like quantum state, which is normalizable
in the physical inner product (\ref{IP4}) {\em and} symmetric
under permutations and reflexions of the $\sigma_p$. The existence of
a normalizable Chern-Simons-like state is in remarkable
contrast to earlier work
on special spatially homogeneous models discussed by Marugan \cite{Maru}.

Finally, we still have to define three further integration contours
$\Sigma^3$ to obtain the complete set of Chern-Simons like solutions.
The construction of the vacuum solutions in appendix A suggests
three new possibilities to create linearly independent
integrals. One of these further integration surfaces can
simply be written in the
form
   
\begin{equation}\label{60}
\Sigma^3_{12}
=\bigl \{ \vec r\,\epsilon\,\mbox{\bf{C}}^3\,|
-\infty<\eta < -1,
\,0<\varphi<2 \pi,\,r\,\epsilon\,
K(1,R) \bigr \}\ \ ,
\end{equation}
\\
while the other two have a very complicated form and may be read off from the
representations (\ref{A16.3}) and (\ref{A16.4}) in appendix A. That
these latter two integration manifolds, despite of their strange
$\eta$-junctions at $-\sigma_1/\sigma_2$ and $-\sigma_2/\sigma_1$,
give rise to analytical solutions of our model even for 
$\Lambda \not= 0$ is shown in appendix C.
However, these solutions do not obey the necessary symmetry properties 
which were derived in section III, and are therefore of no further
interest to us.

One may now finally ask, how the four solutions defined in (\ref{A16.1})
to (\ref{A16.4}) of appendix A are connected to the `asymmetric' solutions
$\Psi'_{\varrho}$ of the diagonal model discussed in \cite{0}. An
investigation
of the semiclassical behavior, together with a discussion of the
singularities on the lines $\sigma_p =\sigma_q,\,p\not= q$, shows
that the solutions $\Psi'_{1},\,\Psi'_{12},\,\Psi'_{123}$ and
$\Psi'_{0123}$ are actually the {\em sums} of these states as indicated
through the choice of their indices, i.e. we have

\begin{equation}\label{61}
\Psi'_{12}=\Psi'_1+\Psi'_2\ \ ,\ \ 
\Psi'_{123}=\Psi'_1+\Psi'_2+\Psi'_3\ \ ,\ \ 
\Psi'_{0123}=\Psi'_0+\Psi'_1+\Psi'_2+\Psi'_3\ \ \ ; 
 \ \ \sigma_3>\sigma_2>\sigma_1\ \ .
\end{equation}
\\
By inverting these relations with respect to $\Psi'_{\varrho}$ one
may further define asymmetric solutions corresponding to those of the
diagonal model, with the same symmetry properties that have
been pointed out in \cite{0}.

\section{Discussion and Conclusion}

In this paper we have examined the transformation connecting the
representations of quantum general relativity in metric variables
and in Ashtekar's variables for the special case of spatially
homogeneous but anisotropic space-times of Bianchi type IX with
{\it non-diagonal} metric tensor. While classically the non-diagonal
case and the diagonal case are equivalent (in the absence of matter)
due to the freedom of gauge-fixing, there is a subtle
difference quantum mechanically, because the steps of gauge-fixing
and quantization, in general, do not commute. This was explained via
a simple example in section III~A. The example also made clear that
gauge-fixing {\it after} quantization is preferable because all
symmetries are then implemented automatically and democratically.
Once the two steps have been completed in this order it is then also
clear {\it a posteriori} how to proceed correctly in the quantization after
the gauge-fixing has been performed on the classical level. Performed
in this way quantization and gauge-fixing of course
commute by construction.

For the non-diagonal Bianchi type IX model we have considered in detail
the transformation of the Chern-Simons state from Ashtekar variables to
metric variables. The Chern-Simons state, including its limit for
vanishing cosmological constant, deserves a thorough study because it is
undoubtedly the most important exact solution of all constraints of
quantum general relativity found up until now. This is so because, unlike
all other exact solutions, it describes a well-defined 
non-degenerate space-time in its classical limit and also
because it makes an obvious connection between quantum general relativity
and topological field theory \cite{Wi,Ka}.
Indeed it is remarkable that the general
Chern-Simons state in Ashtekar variables makes no reference to metric
concepts on the space-like 3-manifold. It should be noted however, that
writing down a physical state in Ashtekar variables does not yet define
it completely, because  `reality conditions' must still be
imposed before a physical interpretation can be attempted. Imposing the
reality conditions is a very nontrivial task and, as our results
indicate, might not have a {\it unique} solution. We circumvent this
problem completely by transforming back to the real metric representation
before applying a gauge-fixing and giving a physical interpretation.

The Ashtekar variables and the densitized inverse triad form canonically
conjugate pairs. Hence, for our Bianchi type IX model the generalized
multidimensional Fourier transformation we discussed in sections IV, V,
VI can be used to transform the Chern-Simons state to the metric
representation. This is a {\it generalized} Fourier transformation because
neither the integration contours, which are here 9-dimensional manifolds,
nor their boundaries are fixed a priori, except for the
condition that partial integration with the Chern-Simons state under the
integral must be permitted without contribution from the boundaries. The
boundaries are therefore determined entirely by the {\it singularities}
of the Chern-Simons state.
On the other hand, for fixed boundaries the different integration
manifolds one can find may be deformed without changing the result, or may be
combined by first running through one integration manifold and then through
other inequivalent ones, leading to linear combinations of the physical
states defined by each integration manifold separately.

In eq.~(\ref{39}),
with the five integration manifolds $\Sigma^3$ given by eqs.~(\ref{59.1}),
(\ref{59.2}), (\ref{60}), (\ref{A16.3}), (\ref{A16.4}), we have
obtained integral representations of five {\it exact} solutions to all
constraints of quantum general relativity and we studied various limits
of these solutions in section V. In the leading semiclassical order
the result  for the non-diagonal and the
diagonal model is the same, as one would expect from the classical
equivalence of both cases. However, differences appear already in the
next to leading order, obtained in the semiclassical expansion of our
results in section V~A. It should be noted that even studying the next to
leading semiclassical order of a physical state in quantum general
relativity is rather nontrivial. It requires to address operator
ordering ambiguities in the Hamiltonian, which we have done in section
II~A, to take proper account of quantum corrections in the Hamiltonian
implied by symmetries, which we have done in section
III~B, and to apply a complete set of gauge-fixing conditions including
a fixing of the time reparametrization symmetry, which we have done in
section III~C. Only then is the next to leading semiclassical order of
a physical state unambiguously defined.

Only two of the five solutions we constructed satisfy the complete permutation
symmetry between the three main axes of the Bianchi IX 3-geometry, which, as
shown in section III~B,
is implied by the quantization of the {\it non-diagonal} model. These are
the states corresponding to the
integration contours (\ref{59.1}) and (\ref{59.2}). Their
semiclassical limits identify them as a wormhole state and a
Hartle-Hawking `no-boundary' state, respectively. It is remarkable that
the {\it same} Chern-Simons state in Ashtekar variables can yield such
diverse states depending on the
choice of the integration manifold. This is a striking example that a physical
state in Ashtekar variables is not yet defined
before further conditions  (here the choice of integration contours) are
specified.

In the Hartle-Hawking proposal for the semiclassical initial condition
of the classical evolution of the Universe it is not required that the
`no-boundary' state is a normalizable vector in a Hilbert space. 
Rather the condition by which it is defined at least semiclassically
is that the Euclidean 4-geometries defined by the semiclassical wavefunction 
filling in the 3-geometries on a space-like slice are regular for $a \to 0$. 
A further requirement for the Hartle-Hawking state is to
give a well-defined probability distribution at the
semiclassical caustic surface where the semiclassical evolution, given by
the wave function, switches from a Riemannian (`Euclidean')
to a Pseudo-Riemannian (`Lorentzian') space-time.
This requirement is met by the Hartle-Hawking `no-boundary'
state obtained here,
and only by this state among the five Chern-Simons-like solutions, as was
shown already in \cite{0}.

The wormhole state, on the other hand, one expects to be a normalizable
vector in a Hilbert space with a well-defined scalar product, in which
other state vectors describe e.g. excited states of the wormhole. Without
going into details, which will be given elsewhere \cite{fut}, we have 
defined such a scalar product in section III~C by gauge-fixing all gauge
symmetries. It turns out that the wormhole state is, in fact, normalizable
in this scalar product, while the `no-boundary state' is not. Thus imposing
either the `no-boundary' condition or normalizability
in the Hilbert space of physical states equipped with a fully gauge-fixed
scalar product, one finds in each case a {\it unique} but mathematically
and physically vastly different state as a metric representation of the
Chern-Simons state.

One may hope that the non-diagonal quantization procedure presented
here might be applicable even to the full
inhomogeneous case, because all constraints are treated quantum
mechanically in a similar way, and a solution to the quantum
constraints is available in terms of the general Chern-Simons
functional.  
Work in this direction is in progress.

\acknowledgements

Support of this work by the Deutsche Forschungsgemeinschaft
through the Sonderforschungsbereich ``Unordnung und gro{\ss}e
Fluktuationen'' is gratefully acknowledged.

\begin{appendix}

\section{The Chern-Simons integral in the limit $\Lambda \to 0$}

In the following we want to examine the possible limits $\Lambda \to 0$
of the Chern-Simons integral given in(\ref{39}). First of all, we
shall briefly recover the wormhole state, which is approached for the 
choice $\Sigma^3=\Sigma^3_{\mbox{{\tiny $W\!H$}}}$ given in (\ref{59.1}).
Using this integration manifold, we may write

\begin{equation}\label{A1}
\Psi'_{\mbox{{\tiny $W\!H$}}} \propto
\int\limits_{-1}^{+1}\,\mbox{d}\eta
\,\int\limits_{0}^{2 \pi}\,\mbox{d}\varphi
\!\!\int\limits_{-\infty+i \varepsilon}^{+\infty+i \varepsilon}
\,\frac{r^2\,\mbox{d}r}{1-r^2}\ \exp \left( \frac{F}{\lambda} \right )\ \ ,
\end{equation}
\\
where we made use of the spherical coordinates in the $(x,y,z)$-space,
which were introduced in (\ref{sc}).
Easy estimates show that for $\Lambda \to 0$ just an infinitesimal
region around $r=0$ contributes to the inner $r$-integral, so we
have the following asymptotic behavior:\footnote{As $\hbar$ is kept
fixed, also $\lambda=\frac{\hbar \Lambda}{3}$, cf. (\ref{31.3}), tends
to zero in the limit $\Lambda \to 0$.} 

\begin{equation}\label{A2}
\Psi'_{\mbox{{\tiny $W\!H$}}}
\ \ \propto^{^{\!\!\!\!\!\!\!\!\!\!\!\!\mbox{{\scriptsize $\lambda\! \to \!0$}}}}
\int\limits_{-1}^{+1}\,\mbox{d}\eta
\,\int\limits_{0}^{2 \pi}\,\mbox{d}\varphi
\int\limits_{-\varepsilon}^{+\varepsilon}
\,\frac{r^2\,\mbox{d}r}{1-r^2}\ \exp \left(\frac{-r^2}{\lambda} \right)
\ \ \sim^{^{\!\!\!\!\!\!\!\!\!\!\!\!\mbox{{\scriptsize $\lambda\! \to \!0$}}}}
4 \pi\, \lambda^{3/2}\int\limits_{-\infty}^{+\infty}\,\mbox{d}\xi\, \xi^2 
e^{-\xi^2}
=2\,(\pi \lambda)^{3/2}\ = \ const.\ \ ,\ \ r=\sqrt{\lambda}\,\xi\ \ .
\end{equation}
\\
According to the transformation rule (\ref{26}) we obviously approach
the wormhole state $\Psi_{\mbox{{\tiny $W\!H$}}}$
in the limit $\Lambda \to 0$, which was defined in (\ref{18}).

Let us now turn to the much more complicated cases, for which the
`wormhole saddle point' $x=y=z=0$ is {\em not} passed through by
the integration surface $\Sigma^{3}$. For such solutions, we shall
try to set not all borders of $\Sigma^3$ at infinity; instead we
shall make use of the existence of two further singularities
of the integrand with respect to $r$, namely those at $r=\pm 1$.
If we integrate into these singularities in a suitable manner, we expect
to create further solutions of our Wheeler-DeWitt equation (\ref{31.1}),
because boundary terms generated by partial integrations will vanish
at these borders. In the following it will be in fact sufficient
to consider only one of the two singularities, say $r=+1$, since
any $r$-integral in the neighbourhood of $r=-1$ can be mapped onto
a region around $r=+1$ by a transform $\eta \to -\eta$ in the
$\eta$-integral. This is due to the fact that the 
$\eta$-,$r$-dependence of the integrand in (\ref{39})
is given in terms of $r^2, r\,\eta$ and $\eta^2$ only, cf. (\ref{expo}) 
below. Thus we will be
interested in integration countours for the $r$-integral, which have
one end point at $r=1$. It turns out that, for a positive cosmological
constant $\Lambda$, there are only two curves
of interest: firstly, we may perform our $r$-integral along the real
axis from $r=1$ to $r=+\infty$; secondly, we can also consider
a pole integral, encircling the singularity at $r=1$ in the mathematically
positive sense. To shorten our notation, we should establish the following
nomenclature: 
a circle with radius $R$, which is centered at $r_0$ and passed through
in the mathematically positive sense will be denoted by $K(r_0,R)$.
Then we can write the integrals of interest in the form

\begin{equation}\label{A3}
\Psi'_{\mbox{{\tiny $C\!S$}}}=
\int \mbox{d}\eta \, \int \mbox{d}\varphi \int_{\cal C}
\,\frac{r^2\,\mbox{d}r}{1-r^2}\ \exp \left [ \frac{G}{\lambda} \right ]\ \ ,
\ \ \mbox{with}
\end{equation}
\begin{equation}\label{expo}
G:=1-r^2+2\,\kappa_3\,r\,\eta+\frac{1}{2}
(\kappa_1^2+\kappa_2^2)+
\frac{1}{2}\,
\frac{(\kappa_1^2+\kappa_2^2)(1+r^2 \eta^2)+4\,\kappa_1\,
\kappa_2\,r\,\eta+r^2\,(\kappa_1^2-\kappa_2^2)\,(\eta^2-1)\cos 2\varphi}
{1-r^2}\ \ ,
\end{equation}
\\
where ${\cal C}$ is one of the two contours $]1,+\infty[$ or $K(1,
R)$, $0<R<2$. The equality sign in (\ref{A3}) indicates that, from now
on, all prefactors of the integral will be taken into account.
To proceed in our calculation of the limit $\Lambda \to 0$, 
we can uniformly estimate those parts of the integral (\ref{A3}),
which lie outside an $\varepsilon$-neighbourhood of $r=1$ by a function,
which vanishes exponentially for $\Lambda \to 0$. Therefore, just an
$\varepsilon$-neighbourhood of $r=1$ contributes to the integral
(\ref{A3}), where $\varepsilon$ may be chosen arbitrarily small. As 
a consequence, all terms of the integrand, which remain regular at $r=1$,
can be substituted by their value taken at $r=1$, and only the $r$-
dependence of singular parts remains to be taken into consideration.  
Thus we arrive at once at

\begin{equation}\label{A4}
\Psi'_{\mbox{{\tiny $C\!S$}}}
\ \ \sim^{^{\!\!\!\!\!\!\!\!\!\!\!\!\mbox{{\scriptsize $\lambda\! \to \!0$}}}}
\frac{1}{2}\,\int \mbox{d}\eta \ e^{\sigma_3 \eta}\,
\int \mbox{d}\varphi
\int_{\mbox{\small{${\cal C}_{\varepsilon}$}}}
 \frac{\mbox{d}r}{1-r}\ \exp \left [
\frac{2}{\lambda}\, (1-r)+\frac{\lambda}{8}\,\frac{a^2-2\, a b \cos 2\varphi
+b^2}{1-r} \right ]\ \ ,
\end{equation}
\\
where we made use of the definitions

\begin{equation}\label{ab}
a:=\frac{\sigma_1+\sigma_2}{2}\,(1+\eta)\ \ ,\ \ 
b:=\frac{\sigma_1-\sigma_2}{2}\,(1-\eta)\ \ .
\end{equation}
\\
Here ${\cal C}_{\varepsilon}$ is either equal to $]1,1+\varepsilon[$ or
given by $K(1,\varepsilon)$. 
After the transformation $\xi=\frac{4}{\lambda}(r-1)$ we
arrive at

\begin{equation}\label{A5}
\Psi'_{\mbox{{\tiny $C\!S$}}}
\ \ \sim^{^{\!\!\!\!\!\!\!\!\!\!\!\!\mbox{{\scriptsize $\lambda\! \to \!0$}}}}
-\frac{1}{2}\,\int \mbox{d}\eta \ e^{\sigma_3 \eta}\,
\int \mbox{d}\varphi
\int_{{\cal C}'}\,\frac{\mbox{d}\xi}{\xi}\,\exp\left [
-\frac{\xi}{2}-\frac{a^2-2\, a b \cos 2 \varphi+b^2}{2\,\xi}
\right ] \ \ ,
\end{equation}
\\
where now ${\cal C}'$ is the positive real axis or $K(0,R)$. The two
possible $\xi$-integrals can now easily be performed, leading to

\begin{equation}\label{A6}
\Psi'_{\mbox{{\tiny $C\!S$}}}
\ \ \sim^{^{\!\!\!\!\!\!\!\!\!\!\!\!\mbox{{\scriptsize $\lambda\! \to \!0$}}}}
\,\int \mbox{d}\eta \ e^{\sigma_3 \eta}\,
\int \mbox{d}\varphi
\ \Biggl\{
\begin{array}{rcl}
-i\,\pi\,I_0 \left [\sqrt{a^2-2\, a b \cos 2 \varphi+b^2}\right ]
&,&{\cal C}'=K(0,R) \\[0.3 cm]
-\ K_0 \left [\sqrt{a^2-2\, a b \cos 2 \varphi+b^2} \right ]
& ,&{\cal C}'=\,]0,+\infty[\ \ .
\end{array}
\end{equation}
\\
Up to now we have taken into acount all freedom in the choice of the
integration contours; any other choice for the $r$-integration contour
would have given a linear combination of the two results given in
(\ref{A6}) or a constant as in (\ref{A2}). While the $\eta$-dependence
of $(\ref{A6})$, hidden in $a$ and $b$, is to complicated to be
integrated out analytically, there are several possibilities to
evaluate the $\varphi$-integral, which become more transparent in the new
variable $u=\cos 2\varphi$. With the abbreviation
$X:=a^2-2\,a b\, u+b^2$ we find

\begin{eqnarray}\label{A9.1}
\int\limits_{-1}^{+1}\frac{\mbox{d}u}{\sqrt{1-u^2}}\,K_0(\sqrt{X})
&=&\pi\,I_0(a)\,K_0(b)\ \ ;\ \ \
b>a>0\ \ ,
\\[.3 cm]
\label{A9.2}
\int\limits_{-1}^{+1}\frac{\mbox{d}u}{\sqrt{1-u^2}}\,I_0(\sqrt{X})
&=&\pi\,I_0(a)\,I_0(b)\ \ ,
\\[.3 cm]\label{A9.3}
\int\limits_{-\infty}^{-1}\frac{\mbox{d}u}{\sqrt{u^2-1}}\,K_0(\sqrt{X})
&=&K_0(a)\,K_0(b)\ \ ;\ \ \
a,b>0\ \ ,
\\[.3 cm]\label{A9.4}
\int\limits_{+1}^{+\infty}\frac{\mbox{d}u}{\sqrt{u^2-1}}\,I_0(\sqrt{X})
&=&I_0(a)\,K_0(b)+K_0(a)\,I_0(b)\ \ ;\ \ \
a,b>0\ \ .
\end{eqnarray}
\\
Obviously, the first two integrals correspond to an integration over
real angles $\varphi$, while the latter two integrals require for 
$\varphi$-values lying on the imaginary axis. The first three integrals
are tabulated in mathematical tables like \cite{40}, but the integral
(\ref{A9.4}) seems to be a new result. Thus we shall shortly comment
on a proof of this formula, which, by the way, also holds for the
other integrals (\ref{A9.1}) to (\ref{A9.3}): The idea is to show that
the integral on the left hand side solves the differential equation
of the modified Bessel functions with index zero with respect to $a$
and $b$. After some partial integrations it turns out that
this is indeed the case; however, we thereby gather a boundary term, which exclusively
vanishes for the boundaries chosen in (\ref{A9.1}) to (\ref{A9.4}).
For theses choices we then know that the integrals must be equal to
a linear combination of the four possible products of $I_0$ and $K_0$
at argument $a$ and $b$. The coefficients of these possible four
contributions can then be determined by evaluating the integrals at
$a=b$, and, if necessary, by an additional investigation of the limits
$a \to 0$ and $b \to 0$.

With the integrals (\ref{A9.1}) to (\ref{A9.4}) we may now perform the
$\varphi$-integral in (\ref{A6})
and get a result of the form

\begin{equation}\label{A10}
\Psi'_{\mbox{{\tiny $C\!S$}}}
\ \ \sim^{^{\!\!\!\!\!\!\!\!\!\!\!\!\mbox{{\scriptsize $\lambda\! \to \!0$}}}}
\int \mbox{d}\eta \ e^{\sigma_3 \eta}\,
Z_0^{(1)}(a)\,Z_0^{(2)}(b)\ \ ,
\end{equation}
\\
with $Z_0^{(1)}$ and $Z_0^{(2)}$ being modified Bessel functions with index
zero. However, not any combination of $Z_0^{(1)}$ and $Z_0^{(2)}$ with
$\eta$-borders at $\pm 1$ or $\pm \infty$ will lead to a solution of the
three equations

\begin{equation}\label{A11}
{\cal F}_p\,\Psi'_{\mbox{{\tiny $C\!S$}}}:=
{\cal Q}_p \big |_{\Lambda=0}\  \Psi'_{\mbox{{\tiny $C\!S$}}}
= \left [
\partial_q\,\partial_r-\partial_p+
\frac{\sigma_q\,\partial_r-\sigma_r\,\partial_q}
{\sigma_q^2-\sigma_r^2} \right ]
\,\Psi'_{\mbox{{\tiny $C\!S$}}}=0\ \ ,
\end{equation}
\\
which necessarily should be the case for the limit $\Lambda \to 0$ of
the Chern-Simons integral because of (\ref{31.2}). The reason for
this is that, due to the complexification of the spherical
coordinates (\ref{sc}), we might have integrated effectively 
over a three dimensional manifold $\Sigma^3$, which produces
boundary terms for partial integrations. For example, there is no
reason, why a surface described by $r\, \epsilon\, ]1,+\infty[,
\,\varphi\, \epsilon\, i${\bf{R}}$,\,\eta\, \epsilon\, ]-1,+1[$ should
lead to an integral solving (\ref{A11}), although this integration
contour leads to a finite integral. It will be a nice explicit check
of solvability to investigate, for which choices of
$Z_0^{(1)}$ and $Z_0^{(2)}$ and for which $\eta$-borders      
$\Psi'_{\mbox{{\tiny $C\!S$}}}$ according to (\ref{A10}) becomes a
solution to the three equations (\ref{A11}):

While the operator ${\cal F}_3$ annihilates the integrand of
(\ref{A10}) for arbitrary values of $\eta$, and therefore
$\Psi'_{\mbox{{\tiny $C\!S$}}}$ for arbitrary borders of the
$\eta$-integral, we have to perform suitable partial integrations
to show the vanishing of ${\cal F}_1\,\Psi'_{\mbox{{\tiny $C\!S$}}}$  
and ${\cal F}_2\,\Psi'_{\mbox{{\tiny $C\!S$}}}$. Consequently,
we get some boundary terms, which will not vanish for all possible
forms of (\ref{A10}). A detailed discussion of these boundary terms
reveals that there are exactly four possiblities to arrive at solutions
of (\ref{A11}). For the adopted ordering of the three $\sigma_p$
these solutions are given by the following
integrals:

\begin{equation}\label{A12.1}
\Psi'_{1}=\int\limits_{-\infty}^{-1}\ \mbox{d}\eta\,
e^{\sigma_3 \eta}\,I_0(a)\,K_0(|b|)\ \ \ ,\ \ \ 
\Psi'_{12}=i\,\pi \int\limits_{-\infty}^{-1}\ \mbox{d}\eta\,
e^{\sigma_3 \eta}\,I_0(a)\,I_0(b)\ \ ,
\end{equation}
\begin{equation}\label{A12.2}
\Psi'_{123}=\int\limits_{-\infty}^{+1}\ \mbox{d}\eta\,
e^{\sigma_3 \eta}\,K_0(|a|)\,I_0(b)\ \ ,\ \ 
\Psi'_{0123}=i\,\pi\,\int\limits_{-1}^{+1}\ \mbox{d}\eta\,
e^{\sigma_3 \eta}\,I_0(a)\,I_0(b)\ \ .
\end{equation}
\\
As an example, we want to discuss the action of the operator
${\cal F}_1$ on the wavefunction $\Psi'_{12}$, which can be
written in the form

\begin{equation}\label{A13}
{\cal F}_1\,\Psi'_{\mbox{{\tiny $C\!S$}}}=
\frac{1}{2}\,\left \lbrack
\frac{e^{\sigma_3 \eta}\,
(\eta^2-1)}
{\sigma_2^2-\sigma_3^2}\biggl (
\sigma_3\,\left [I_1(a)\,I_0(b)+I_0(a)\,I_1(b) \right ]  
-\sigma_2\,\left [I_1(a)\,I_1(b)+I_0(a)\,I_0(b) \right ] \biggr )
\right \rbrack_{\eta \to -\infty}^{\eta=-1}\ \ .
\end{equation}
\\
That this is indeed equal to zero is easily seen for $\eta=-1$, and,
after some asymptotic expansions of the Bessel functions, also for
$\eta \to -\infty$. Similar results hold for the other solutions, where
one of the two Bessel functions might be substituted by $K_0$ or
$-K_1$. In such cases the vanishing of the boundary terms at 
$\eta=\pm 1$ becomes a nontrivial question, because the $K_{\nu}$-
Bessel function might become singular at these points. A detailed 
investigation shows that there are only
the four possibiblities given in (\ref{A12.1}) and (\ref{A12.2}) to
avoid contributions from these boundaries. 

The question which remains to be answered at this point is which
integration contours for the $\varphi$-integral give rise to the
integrands in (\ref{A12.1}) and (\ref{A12.2}). For the wavefunctions
$\Psi'_{12}$ and $\Psi'_{0123}$ the answer is easily found with help
of the integral (\ref{A9.2}), and we find the explicit representations:

\begin{equation}\label{A16.1}
\Psi'_{12}=-\frac{1}{2\,\pi}\,\lim_{\Lambda \to 0}\ \,
\int\limits_{-\infty}^{-1}\mbox{d}\eta\,\int\limits_{0}^{2\,\pi}
\mbox{d}\varphi\,\int_{K(1,R)} \frac{r^2\,\mbox{d}r}{1-r^2}\ 
\exp\left [\frac{G}{\lambda} \right ]\ \ ,
\end{equation}
\begin{equation}\label{A16.2}
\Psi'_{0123}=-\frac{1}{2\,\pi}\,\lim_{\Lambda \to 0}\ \,
\int\limits_{-1}^{+1}\mbox{d}\eta\,\int\limits_{0}^{2\,\pi}
\mbox{d}\varphi\,\int_{K(1,R)} \frac{r^2\,\mbox{d}r}{1-r^2}\ 
\exp\left [\frac{G}{\lambda} \right ]\ \ ,
\end{equation}
\\
where the Radius $R$ of the $r$-integral has to be chosen with $R<2$.
However, for the other two wavefunctions the answer is more difficult,
because, according to (\ref{A9.1}) to (\ref{A9.4}), there is no easy
possibility to generate their integrands uniformly for all $\eta$-values
of their integration regime. So we have to cut the $\eta$-integral
artificially at those $\eta$-values, for which we get $a=\pm b$, which
actually happens at $\eta=-\sigma_2/\sigma_1<-1$
and $\eta=-\sigma_1/\sigma_2>-1$. Suitable linear combinations
of the results (\ref{A9.1}) and (\ref{A9.4}) then reveal the following
integral representations:

\begin{equation}\label{A16.3}
\Psi'_{1}=-\frac{1}{2\,\pi}\,\lim_{\Lambda \to 0}\ \,\left (\,
\int\limits_{-\infty}^{-1}\mbox{d}\eta\,\int\limits_{0}^{2\,\pi}
\mbox{d}\varphi\,\int\limits_{1}^{+\infty}
-2\,
\int\limits_{-\infty}^{-\sigma_2/\sigma_1}\mbox{d}\eta
\,\int\limits_{0}^{2\,\pi}
\mbox{d}\varphi\,\int\limits_{1}^{+\infty}
-\,
\int\limits_{-\infty}^{-\sigma_2/\sigma_1}\mbox{d}\eta
\,\int\limits_{-i\,\infty}^{+i\,\infty}
\mbox{d}\varphi\,\int_{K(1,R)} 
\right )
\frac{r^2\,\mbox{d}r}{1-r^2}\ 
\exp\left [\frac{G}{\lambda} \right ]\ \ ,
\end{equation}
\begin{equation}\label{A16.4}
\Psi'_{123}=-\frac{1}{2\,\pi}\,\lim_{\Lambda \to 0}\ \,\left (\,
\int\limits_{-\infty}^{+1}\mbox{d}\eta\,\int\limits_{0}^{2\,\pi}
\mbox{d}\varphi\,\int\limits_{1}^{+\infty} 
-2\,
\int\limits_{-\sigma_2/\sigma_1}^{-\sigma_1/\sigma_2}
\mbox{d}\eta
\,\int\limits_{0}^{2\,\pi}
\mbox{d}\varphi\,\int\limits_{1}^{+\infty}
-\,
\int\limits_{-\sigma_2/\sigma_1}^{-\sigma_1/\sigma_2}
\mbox{d}\eta
\,\int\limits_{-i\,\infty}^{+i\,\infty}
\mbox{d}\varphi\,\int_{K(1,R)} 
\right )
\frac{r^2 \mbox{d}r}{1-r^2}\ 
\exp\left [\frac{G}{\lambda} \right ]\ \ .
\end{equation}
\\
Considering the representations (\ref{A16.1}) to (\ref{A16.4}) it is
now possible to read off the integration manifolds $\Sigma^3$ which
lead to solutions of the Wheeler-DeWitt equation even for a non
vanishing cosmological constant $\Lambda$. Although the representations
(\ref{A16.3}) and (\ref{A16.4}) look very strange because of the
$\eta$-junctions at $-\sigma_2/\sigma_1$ and 
$-\sigma_1/\sigma_2$, they lead indeed to differentiable
solutions. Since this is a non trivial claim, it will be proven in
appendix C, where we show that the integrand of the outer $\eta$-integral
is a continous and differentiable function at the junction points, even
for $\Lambda \not= 0$.

\section{Alternative integral representations for the vacuum solutions}

We will now bring the one dimensional integral representations found
in appendix A into a new, unified form, which will display nicely the
symmetries of these solutions. The motivation of this new representation
arises from the fact that one of the $\eta$-integrals, namely $\Psi'_{0123}$
given in (\ref{A12.1}), turns out to be a convolution integral, and may
therefore be simplified with the aid of Laplace's convolution theorem.
To become more precise, let us write

\begin{equation}\label{B1}
\Psi'_{0123}=2\pi i\,\int\limits_{0}^{1}\,\mbox{d}\tau
\,e^{-\sigma_3 \tau}\,I_0 \bigl [(\sigma_1-\sigma_2)\tau \bigr ]\,
e^{\sigma_3(1-\tau)}\,I_0 \bigl [(\sigma_1+\sigma_2)(1-\tau) \bigr ]\ \ ,
\end{equation}
\\
where we have substituted $\tau=\frac{1-\eta}{2}$ in (\ref{A12.2}). If we denote

\begin{equation}\label{B2}
g(\alpha,\beta;t):=e^{\alpha t}\,I_0[\beta t]\ \ ,
\end{equation}
\\
we may also write

\begin{equation}\label{B3}
\Psi'_{0123}=2\pi i\ g(-\sigma_3,\sigma_2-\sigma_1;t)\star
g(\sigma_3,\sigma_2+\sigma_1;t) \, \big |_{t=1}\ \ ,
\end{equation}
\\
i.e. (\ref{B1}) is a convolution integral, evaluated at $t=1$. Let us 
recall Laplace's convolution theorem, which states

\begin{equation}\label{B4}
{\cal L}[f_1 \star f_2](s)={\cal L}[f_1](s)\cdot{\cal L}[f_2](s)\ \ ,
\ \ \ \mbox{with}\ \ \ {\cal L}[f](s):=\int\limits_{0}^{\infty}\mbox{d}t\,
f(t)\ \ .
\end{equation}
\\
Using the inverse Laplace transform

\begin{equation}\label{B6}
f(t)=\frac{1}{2 \pi i}\,\int\limits_{\hat s-i\,\infty}^{\hat s+i\,\infty}
 \mbox{d}s\,e^{s t}\,{\cal L}[f](s)\ \ ,
\end{equation}
\\
where $\hat s$ has to lie right to all singularities of the integrand
in (\ref{B6}), Laplace's convolution theorem can also be written in
the form

\begin{equation}\label{B7}
(f_1 \star f_2)\,(t)=\frac{1}{2 \pi i}\,
\int\limits_{\hat s-i\,\infty}^{\hat s+i\,\infty}
 \mbox{d}s\,e^{s t}\,{\cal L}[f_1](s)\cdot{\cal L}[f_2](s)\ \ .
\end{equation}
\\
With aid of the Laplace transform of $g$ defined in (\ref{B2}), 

\begin{equation}\label{B8}
{\cal L}[g](s)=\int\limits_{0}^{\infty} \mbox{d}t\,e^{(\alpha-s)t}\,
I_0[\beta t]=[(\alpha-s)^2-\beta^2]^{-\frac{1}{2}}\ \ ,
\end{equation}
\\
we arrive at once at the identity

\begin{equation}\label{B9}
\Psi'_{0123}=\int\limits_{\hat s-i\,\infty}^{\hat s+i\,\infty}
\mbox{d}s\,e^s\,\prod_{\nu=0}^{3}\,(s-s_{\nu})^{-\frac{1}{2}}\ \ ,
\end{equation}
\\
if we employ (\ref{B7}) with $t=1$. Here we have introduced the four 
quantities $s_{\nu}$ via (\ref{sss}),
which are ordered according to $s_1 < s_2 < s_3 < s_0$ for 
$\sigma_1< \sigma_2 < \sigma_3$. 
Since the quantity $\hat s$ occuring
in (\ref{B9}) must be placed right to all the $s_{\nu}$, it must be
chosen as $\hat s > s_0$. With (\ref{B9}) we have reached a very nice
representation for $\Psi'_{0123}$, which immediately shows up the 
symmetry with respect to arbitrary permutations of the $s_{\nu}$:
As the integrand obviously obeys this symmetry, and the integration
curve, which is placed right to all the $s_{\nu}$ can be chosen to be
the same after 
two of the $s_{\nu}$ have been permuted, the integral must be invariant under
such permutations. Translated to the $\sigma_p$, we firstly 
have a symmetry under
permutations of the $\sigma_p$ themselves. Secondly, the additional symmetry
under permutations
of the form $s_{0} \leftrightarrow s_p$ implies that the wavefunction
$\Psi'_{0123}$ is invariant under a reflexion $\sigma_1 \to -\sigma_1$,
$\sigma_2 \to -\sigma_2,\sigma_3 \to \sigma_3$ and cyclic permutations
thereof. Thus we directly recover the symmetry properties of
$\Psi'_{0123}$, which we already claimed in section VI.

Regarding the result (\ref{B9}) in more detail, one may now ask, if there
could be further integration curves, apart from the one used in
(\ref{B9}), leading also to vacuum solutions. Investigating the action
of the operators ${\cal F}_p$ defined in (\ref{A11}) we find the following
result

\begin{equation}\label{B11}
{\cal F}_3\ \left [ \int \mbox{d}s\,e^s \,
\prod_{\nu=0}^{3}\,(s-s_{\nu})^{-\frac{1}{2}} \right ]
=\frac{1}{2}\,
\int \mbox{d}s\,\frac{\mbox{d}}{\mbox{d}s} \left [
\left (
-\frac{1}{s-s_0}+\frac{1}{s-s_1}+\frac{1}{s-s_2}-\frac{1}{s-s_3}
\right )
\ e^s\,\prod_{\nu=0}^{3}\,(s-s_{\nu})^{-\frac{1}{2}}
\right ]\ \ ,
\end{equation}
\\
and similar results hold for the action of the operators
${\cal F}_1$ and ${\cal F}_2$ on the $s$-integral. Thus we conclude
that all four integration curves shown in figure \ref{fig1} lead
to solutions of ${\cal F}_p\,\Psi'=0$, because the boundary term
(\ref{B11}) vanishes at the end points of theses curves. However,
up to now it remains unclear, how these vacuum solutions are connected
to those which were found in appendix A. In the following, we shall be
interested in a direct transformation between the two representations,
which is rather tricky and therefore should be presented in detail at
least for one of the remaining solutions. In our calculations, we will
need the following two integrals:

\begin{equation}\label{B12}
e^{\alpha t}\, I_0[\beta t]=
\frac{1}{2 \pi i}\,\int\limits_{\hat s-i\,\infty}^{\hat s+i\,\infty}
 \frac{\mbox{d}s\,e^{s t}}{\sqrt{(\alpha-s)^2-\beta^2}}\ \ , \ \
\hat s> \alpha \pm \beta\ \ ,
\end{equation}
\begin{equation}\label{B13}
\int\limits_{-\infty}^{+\infty} \mbox{d}t\,e^{\alpha t}\,
K_0[\beta\,|t|]=\frac{\pi}{\sqrt{\beta^2-\alpha^2}}\ \ .
\end{equation}
\\
Let us now choose as an example the transformation
of the wavefunction $\Psi'_1$:

\begin{eqnarray}\label{B14}
\Psi'_1&=&2\,\int\limits_{0}^{\infty} \mbox{d}\tau\,e^{-\sigma_3(\tau+1)}
\,K_0 \bigl [(\sigma_2-\sigma_1)(\tau+1) \bigr ]\,
e^{-\sigma_3 \tau}\,I_0 \bigl [(\sigma_1+\sigma_2) \tau \bigr ]\ \ ,
\ \ \tau=-\frac{1+\eta}{2}\nonumber\\[.3 cm]
&\mbox{\footnotesize{(\ref{B12})}}&\nonumber\\[-.5 cm]
&=&\frac{1}{i \pi}\,\int\limits_{0}^{\infty} \mbox{d}\tau\,e^{-\sigma_3(\tau+1)}
\,K_0 \bigl [(\sigma_2-\sigma_1)(\tau+1) \bigr ]\,
\int\limits_{\hat s-i\,\infty}^{\hat s+i\,\infty}
\frac{\mbox{d}s'\,e^{s' \tau}}{\sqrt{(s'+s_0)(s'+s_3)}}\ \ ,
\ \ \hat s>-s_0,-s_3\nonumber\\[.3 cm]
&_{\tau'=1+\tau}&\nonumber\\[-.5 cm]
&=&
\frac{1}{i \pi}\,\int\limits_{\hat s-i\,\infty}^{\hat s+i\,\infty}
\mbox{d}s\,e^s\,\bigl [ (s-s_0)(s-s_3) \bigr ]^{-\frac{1}{2}}\ 
\int\limits_{1}^{\infty}\mbox{d}\tau'\,
e^{-(\sigma_3+s)\tau'}\,K_0 \bigl [(\sigma_2-\sigma_1)|\tau'| \bigr]\ \ ,
\ \ \hat s<s_0,s_3\ \ .\nonumber\\[-.7 cm]
&_{s=-s'}&
\end{eqnarray}
\\
To allow for the interchange of the $\tau'$- and the $s$-integral 
performed in the last two lines, we further have to make sure that 
$\hat s >s_1$, otherwise the $\tau'$-integral in the last line would
not exist. If we further require $\hat s< s_2$, the $\tau'$-integral will
exist even for an extension of the $\tau'$-integral to $-\infty$.
Let us consider the additional contribution which we would get in case
of this extension:

\begin{eqnarray}\label{B15}
&&\int\limits_{\hat s-i\,\infty}^{\hat s+i\,\infty}
\mbox{d}s\,e^s\,\bigl [ (s-s_0)(s-s_3) \bigr ]^{-\frac{1}{2}}\ 
\int\limits_{-\infty}^{+1}
\mbox{d}\tau'\,
e^{-(\sigma_3+s)\tau'}\,K_0 \bigl [(\sigma_2-\sigma_1)|\tau'| \bigr]
\nonumber\\[.3 cm]
&=&
\int\limits_{-\infty}^{+1}
\mbox{d}\tau'\,
e^{-\sigma_3 \tau'}\,K_0 \bigl [(\sigma_2-\sigma_1)|\tau'| \bigr]\,
\underbrace{
\int\limits_{\hat s-i\,\infty}^{\hat s+i\,\infty}
\mbox{d}s\,
e^{s(\overbrace{1-\tau'}^{>0})}\,\bigl [ (s-s_0)(s-s_3) \bigr ]^{-\frac{1}{2}}
}_0
\ =\ 0\ \ .
\end{eqnarray}
\\
Here the last $s$-integral vanishes, because,
without meeting any singularity of the integrand,
the integration contour can
be deformed to the negative real axis,
where the $s$-integrand vanishes exponentially.
Thus, we still
have a representation of $\Psi'_1$, if we extend the
$\tau'$-integral in the last line of (\ref{B14}) to the whole real axis.
After employing the formula (\ref{B13}) we arrive at

\begin{eqnarray}\label{B16}
\Psi'_1&=&\frac{1}{i}\,
\int\limits_{\hat s-i\,\infty}^{\hat s+i\,\infty}
\mbox{d}s\,\bigl [ (s-s_0)(s-s_3) \bigr ]^{-\frac{1}{2}}\,
\bigl [ (s_2-s)(s-s_1) \bigr ]^{-\frac{1}{2}}
\nonumber\\[.3 cm]
&=&\int\limits_{\hat s-i\,\infty}^{\hat s+i\,\infty}
\mbox{d}s\,e^s \,
\prod_{\nu=0}^{3}\,(s-s_{\nu})^{-\frac{1}{2}}\ \ ,
\ \ s_1<\hat s <s_2\ \ .
\end{eqnarray}
\\
Similar calculations reveal the remaining two identities

\begin{equation}\label{B17}
\Psi'_{12}=
\int\limits_{\hat s-i\,\infty}^{\hat s+i\,\infty}
\mbox{d}s\,e^s \,
\prod_{\nu=0}^{3}\,(s-s_{\nu})^{-\frac{1}{2}}\ \ ,
\ \ s_2<\hat s <s_3\ \ ,
\end{equation}
\begin{equation}\label{B18}
\Psi'_{123}=
\int\limits_{\hat s-i\,\infty}^{\hat s+i\,\infty}
\mbox{d}s\,e^s \,
\prod_{\nu=0}^{3}\,(s-s_{\nu})^{-\frac{1}{2}}\ \ ,
\ \ s_3<\hat s <s_0\ \ .
\end{equation}
It is now easy to comment on the symmetries of the wavefunctions
(\ref{B16}) to (\ref{B18}):
While $\Psi'_{123}$ obviously is symmetric under permutations of the
$s_p$, and therefore under permutations of the $\sigma_p$, 
it is not symmetric under a
permutation $s_p \leftrightarrow s_0$ as the wavefunction $\Psi'_{0123}$.
The vacuum state $\Psi'_{12}$ is only symmetric under the permutation
$s_1 \leftrightarrow s_2$, and, consequently, under $\sigma_1
\leftrightarrow \sigma_2$; the solution $\Psi'_1$ has no symmetry of this
form at all.

Finally, we shall be interested in the limit $\hbar \to 0$ of our
vacuum solutions, where we should remind the reader that 
the $\hbar$-dependence is hidden in the $s_{\nu}$
via the $\sigma_p$, which depend on $\hbar$ according to (\ref{sig}).
As an example, let us compute the limit $\hbar \to 0$ of the
wavefunction $\Psi'_1$:
First of all, the integration contour ${\cal C}_1$ can be deformed
to the negative real axis, giving

\begin{equation}\label{B18.0}
\Psi'_1=-2\,\int\limits_{-\infty}^{s_1}
\mbox{d}s\,e^s \,
\prod_{\nu=0}^{3}\,(s-s_{\nu})^{-\frac{1}{2}}\ \ .
\end{equation}
\\
Substituting $\bar s_{\nu}=\hbar s_{\nu}$ and $\bar \xi=\hbar(s_1-s)$ we
get

\begin{equation}\label{B18.1}
\Psi'_1=-2\, \hbar\,e^{s_1}\,\int\limits_{0}^{\infty}
\frac{\mbox{d}\bar \xi}{\sqrt{\bar \xi}}\,
e^{-\bar \xi /\hbar}\,
\bigl [
(\bar \xi-{\bar s}_1+{\bar s}_2)(\bar \xi-{\bar s}_1+
{\bar s}_0)(\bar \xi-{\bar s}_1+{\bar s}_3)
 \bigr ]^{-\frac{1}{2}}\ \ .
\end{equation}
\\
In the limit $\hbar \to 0$ just an infinitesimal neighbourhood around
$\bar \xi=0$ contributes to the $\bar \xi$-integral, thus we may expand

\begin{equation}\label{B18.2}
\Psi'_1\ 
\, \sim^{^{\!\!\!\!\!\!\!\!\!\!\!
\mbox{{\scriptsize $\hbar\! \to \!0$}}}}
\,
-2\, \hbar\,e^{s_1}\,\int\limits_{0}^{\varepsilon}
\frac{\mbox{d}\bar \xi}{\sqrt{\bar \xi}}\,
e^{-\bar \xi /\hbar}\,
\bigl [
({\bar s}_1-{\bar s}_2)({\bar s}_1-{\bar s}_0)({\bar s}_3-{\bar s}_1)
 \bigr ]^{-\frac{1}{2}}\ \ ,
\end{equation}
\\
where we have substituted $\bar \xi=0$ in those parts of the integrand,
which remain regular for $\hbar \to 0,\bar \xi \to 0$, since we may choose
$\varepsilon$ arbitrarily small. In the new variable $\xi=\bar \xi/
\hbar$, we finally arrive at

\begin{equation}\label{B18.3}
\lim_{\hbar \to 0}\,\Psi'_1\, 
=\frac{- e^{s_1}}{\sqrt{2\,(\sigma_2-\sigma_1)(\sigma_2+\sigma_3)
(\sigma_3-\sigma_1)}}\ 
\int\limits_{0}^{\infty}\,\frac{\mbox{d}\xi}{\sqrt{\xi}}\,e^{-\xi}
=-\sqrt{\frac{\pi}{2}}\,\frac{e^{\sigma_1-\sigma_2-\sigma_3}
}{\sqrt{(\sigma_2-\sigma_1)(\sigma_2+\sigma_3)
(\sigma_3-\sigma_1)}}\ \ .
\end{equation}
\\
The limit $\hbar \to 0$ of the other vacuum solutions may be calculated
analogously. As a result, we obtain the same asymptotic behavior as in
the limit $\hbar \to 0, \Lambda \to 0$, which we discussed
in (\ref{48}) of section V, i.e. the result is independent of the order
in which these two limits are taken.

\section{Continuity and differentiability of the 
integrand on the asymmetric integration surfaces}

We finally want to show that the two integrals (\ref{A16.3}) and 
(\ref{A16.4}) given in appendix A give analytical solutions of the
Wheeler DeWitt equation (\ref{31.1}) not only for $\Lambda \to 0$, but
also for $\Lambda \not= 0$. To show this we have to prove that the
effective integrand of the $\eta$-integral is continous and
differentiable at the junctions $\eta=-\sigma_1/\sigma_2$ and
$\eta=-\sigma_2/\sigma_1$; the Wheeler DeWitt operator
would otherwise produce boundary terms when acting on the wavefunctions
$\Psi'_{1}$ and $\Psi'_{123}$, and they could not be solutions.
In the following, we will restrict ouselves to the solution $\Psi'_1$
in the case $\Lambda > 0$ defined in (\ref{A16.4}), which can be written
in the form:

\begin{equation}\label{C1}
2 \pi\,\Psi'_1=-
\int\limits_{-\infty}^{-1}\mbox{d}\eta \,\int\limits_{0}^{2 \pi}
\mbox{d}\varphi\,\int\limits_{1}^{\infty}\frac{r^2\,\mbox{d}r}{1-r^2}\,
\exp \left [\frac{G}{\lambda} \right ]+
\int\limits_{-\infty}^{-\sigma_2/\sigma_1}
\!\!\!\!\mbox{d}\eta \left (
2 \,\int\limits_{0}^{2 \pi}
\mbox{d}\varphi\,\int\limits_{1}^{\infty}+
\int\limits_{-i\, \infty}^{+i\,\infty}\!\! \mbox{d}\varphi\,
\int_{K(1,R)} \right )
\frac{r^2\,\mbox{d}r}{1-r^2}\,
\exp \left [\frac{G}{\lambda} \right ]\ \ .
\end{equation}
\\
With the abbreviations $P$ and $Q$ defined via
   
\begin{equation}\label{PQ}
P:=\int\limits_{0}^{2 \pi}
\mbox{d}\varphi\,\int\limits_{1}^{\infty}\frac{r^2\,\mbox{d}r}{1-r^2}\,
\exp \left [\frac{G}{\lambda} \right ]\ \ ,\ \ 
Q:=\frac{1}{2}\,\int\limits_{-i\, \infty}^{+i\,\infty}\!\! \mbox{d}\varphi\,
\int_{K(1,R)}\frac{r^2\,\mbox{d}r}{1-r^2}\,
\exp \left [\frac{G}{\lambda} \right ]\ \ , 
\end{equation}
\\
we can rewrite (\ref{C1}) as a piecewise defined $\eta$-integral:   
  
\begin{equation}\label{C1+}
2 \pi\,\Psi'_1=\int\limits_{-\infty}^{-1}\mbox{d}\eta\ 
\biggl\{
\begin{array}{ccl}
P+2\,Q&,&\eta<-\sigma_2/\sigma_1\\[.2 cm]
-P&,&\eta>-\sigma_2/\sigma_1
\end{array}\ \ \ .
\end{equation}
\\
To show the continuity and differentiability of the integrand with
respect to $\eta$ at the $\eta$-junction we then have to show the
following two properties of $P$ and $Q$:   

\begin{equation}\label{C2}
\left (P+Q \right ) \big |_{\eta=-\sigma_2/\sigma_1}=0\ \ ,\ \ 
\frac{\mbox{d}}{\mbox{d}\eta}\,
\left (P+Q \right ) \big |_{\eta=-\sigma_2/\sigma_1}=0\ \  .
\end{equation}
\\
To proceed in this direction let us try to evaluate the 
$\varphi$-integrals in $P$ and $Q$. The $\varphi$-integral for $P$
can be interchanged with the $r$-integral and is then 
peformed easily, leading to
   
\begin{equation}\label{C4}
P=2 \pi\,\int\limits_{1}^{\infty}\frac{r^2\,\mbox{d}r}{1-r^2}\,
I_0 \left [\frac{\lambda}{2}\,a b\,\frac{r^2}{1-r^2} \right ]\,
\exp\left [\frac{G'}{\lambda}
\right ]\ \ ,
\end{equation}
\begin{equation}\label{C4+}
\mbox{with}\ \ 
G':=1-r^2+2\,\kappa_3 r \eta+\frac{1}{2}\,
(\kappa_1^2+\kappa_2^2)\,(1-\eta^2)+\frac{\lambda^2}{4}\,\frac{
a^2+b^2}{1-r^2}-\frac{2\,\kappa_1\kappa_2 \eta}{1+r}\ \ .
\end{equation}
\\
The $Q$-integral is more complicated, because it is
not allowed to interchange the $r$-and the 
$\varphi$-integral which have to be performed there: after such
an exchange
the $\varphi$-integral would not exist. However, we are free to open
the $r$-integration contour $K(1,R)$ at $r \to +\infty$, and the 
resulting integration curve can be deformed into a line integral 
$r_0-i\,${\bf{R}}, where we can choose $-1<r_0<0$. For this 
$r$-integration contour we may then interchange the $r$- and the
$\varphi$-integral, because the $\varphi$-integral exists for any
$r$-value along the contour. Thus we can write
     
\begin{equation}\label{C5}
2\,Q=\int\limits_{r_0+i\,\infty}^{r_0-i\,\infty}
\frac{r^2\,\mbox{d}r}{1-r^2}\,
\int\limits_{-i\,\infty}^{+i\,\infty} \mbox{d}\varphi\,\exp \left[
\frac{G}{\lambda} \right ]\ 
=\ 2\,i\,\int\limits_{r_0+i\,\infty}^{r_0-i\,\infty}
\frac{r^2\,\mbox{d}r}{1-r^2}\,
K_0 \left [\frac{\lambda}{2}\,a b\,\frac{r^2}{1-r^2} \right ]\,
\exp\left [\frac{G'}{\lambda}
\right ]\ \ , 
\end{equation}
\\
and in the last line the $r$-integral can be deformed back to the
positive real axis, giving an integral around the cut for $r>1$,
which is generated by the $K_0$-Bessel function. Let us define this
cut integral explicitely in terms of the contour ${\cal C}_{cut}$ via
   
\begin{equation}\label{C6}
{\cal C}_{cut}:=K(1,R)\oplus \{r-i\,0\,|\,1+R<r<\infty \}
\ominus\{r+i\,0\, |\,1+R<r<\infty \}\ \ .
\end{equation}
\\
Obviously, ${\cal C}_{cut}$ is a superposition of a pole integral
and two line integrals along the cut, where one of these line integrals
is performed in the upper half plane, while the other one has to
be evaluated in the lower half plane. The integral over the
integration curve (\ref{C6})
is independent of the Radius $R$, as long as this ranges between $R=2$ and
$R=0$. Using the formulas

\begin{equation}\label{C7}
K_0(-t\pm i\,0)=\mp\, i \pi I_0(t)+K_0(t)\ \ ,\ \ t>0\ \ ,
\end{equation}
\\
we can evaluate the two line integrals which are contained in (\ref{C6}),
where it turns out that the $K_0$-contributions cancel, while we get
the $I_0$-contributions twice. Thus we can rewrite (\ref{C5}) as

\begin{equation}\label{C8}
2\,Q=-4\,\pi\,\int\limits_{1+R}^{\infty}
\frac{r^2\,\mbox{d}r}{1-r^2}\,
I_0 \left [\frac{\lambda}{2}\,a b\,\frac{r^2}{1-r^2} \right ]\,
\exp\left [\frac{G'}{\lambda}
\right ]
+2\,i\,\int_{K(1,R)}
\frac{r^2\,\mbox{d}r}{1-r^2}\,
K_0 \left [\frac{\lambda}{2}\,a b\,\frac{r^2}{1-r^2} \right ]\,
\exp\left [\frac{G'}{\lambda}
\right ]\ \ .
\end{equation}
\\
If we now consider the limit $R \to 0$ in (\ref{C8}), the first
integral takes the form of the $P$-integral, cf. (\ref{C4}), and we get

\begin{equation}\label{C9}
2\,Q=-2\,P+2\,i\,\lim_{R \to 0}\,
\int_{K(1,R)}
\frac{r^2\,\mbox{d}r}{1-r^2}\,
K_0 \left [\frac{\lambda}{2}\,a b\,\frac{r^2}{1-r^2} \right ]\,
\exp\left [\frac{G'}{\lambda}
\right ]\ \ .
\end{equation}
\\
Comparing this result with the desired claim (\ref{C2}), all
that remains to be shown is that the pole integral in (\ref{C9}) and
its $\eta$-derivative vanish in the limit $R \to 0$. In this limit
$R \to 0$ the only contribution to the integral arises from an
infinitesimal region around $r=1$, so we can expand the 
$K_0$-Bessel function for large arguments; furthermore,
parts of the integrand,
which remain regular for $r \to 1$, may be substituted by their
values taken at $r=1$. We then find for the pole integral

\begin{equation}\label{C10}
\int_{K(1,R)}
\frac{r^2\,\mbox{d}r}{1-r^2}\,
K_0 \left [\frac{\lambda}{2}\,a b\,\frac{r^2}{1-r^2} \right ]\,
\exp\left [\frac{G'}{\lambda}
\right ]
\ \ \ \propto^{^{\!\!\!\!\!\!\!\!\!\!\!
\mbox{{\scriptsize $R\! \to \!0$}}}}
\ \int_{K(1,R)}\frac{\mbox{d}r}{\sqrt{1-r}}\,\exp \left [
\frac{\lambda}{8}\,\frac{(a-b)^2}{1-r} \right ]\ \ ,
\end{equation}
\\
where we have only taken into account the $r$-dependent parts of
the integral, while a $\sigma_p$-dependent prefactor has been omitted.
If we now consider the right hand side of eq. (\ref{C10})
at the $\eta$-junction $\eta=-\sigma_2/\sigma_1$, $a$ and $b$ become
coincident. Thus the essential singularity of the integrand disappears,
and all that remains is an integrable square-root singularity, for
which the pole integral vanishes in the limit $R \to 0$.
Similar arguments show that also the
$\eta$-derivative of the pole integral in (\ref{C9}) vanishes for 
$R \to 0$, so, after all, we have proven our claim (\ref{C2}) and,
therefore, the continuity and differentiability of the $\eta$-integrand
in (\ref{C1}). Analgous calculations may be performed for the second
non trivial integration surface $\Sigma^3_{123}$, where we have to
deal with two different $\eta$-junctions.

\begin{equation}\label{}
\end{equation}

\end{appendix}

\end{document}